\documentclass[aps,twocolumn,superscriptaddress,showpacs,floatfix,preprintnumbers]{revtex4-1}

\usepackage{graphicx} 
\usepackage{float}
\usepackage{dcolumn}
\usepackage{bm}
\usepackage{amsmath,xcolor,comment}

\newcommand{\bA}{\bar{A}}
\newcommand{\hM}{\hat{M}}

\usepackage{float}
\usepackage{amsmath,amssymb}

\usepackage{cleveref}

\begin{document}

\title{On instabilities of perturbations in some homogeneous color-electric and -magnetic backgrounds in $SU(2)$ gauge theory.}

\author{Divyarani C. Geetha,}
\affiliation{Faculty of Science and Technology, University of Stavanger, 4036 Stavanger, Norway}
\author{Anders Tranberg}
\affiliation{Faculty of Science and Technology, University of Stavanger, 4036 Stavanger, Norway}

\date{March 2024}


\begin{abstract}
We consider the instabilities of field perturbations around a homogeneous background color-electric and/or -magnetic field in SU(2) pure gauge theory. We investigate a number of distinct cases of background magnetic and electric fields, and compute the dispersion relations in the linearised theory, identifying stable and unstable momentum modes. In the case of a net homogenous non-abelian $B$-field, we compute the non-linear (quadratic and cubic) corrections to the equation of motion, and quantify to what extent the instabilities are tempered by these non-linearities. 
\end{abstract}

\maketitle

\section{Introduction}
\label{sec:Intro}
The thermalisation process of non-abelian gauge fields has been the subject of intense study, in particular in the context of heavy ion collisions (HIC) (see \cite{Mrow3} for a comprehensive review). In the initial stages of the collision, anisotropic particle distributions along and perpendicular to the beam line may create large anisotropic classical fields \cite{lappi1,lappi2, lappi3,epelbaum1, epelbaum2}. In this background, which is not necessarily homogeneous, perturbations may under certain conditions grow (semi-)exponentially. These (plasma) instabilities may result in long-wavelength fluctuations with very large occupation numbers, that may impact the equilibration process as well \cite{Baier,Mrow1,Mrow2,Mrow3}. It turns out that fast hydrodynamization may likely be understood from kinetic theory with input from perturbation theory \cite{kurk1,kurk2,kin1,kin2,kin3,kin4,schlichting1,schlichting2,schlichting3}, but a thorough understanding of also the non-perturbative dynamics is still of great interest.

Instabilities in anisotropic backgrounds are well-known from both abelian and non-abelian gauge theory \cite{Randrup}. While a U(1) theory produces linear field equations of motion for the gauge field instabilities, which could then in principle grow very large, for QCD the equations of motion are non-linear and the instabilities could turn out to be short-lived and irrelevant. The abelianization phenomenon \cite{Arnold} implies that there will may be directions in field space, where non-linear contributions vanish, and where the instability could continue unhindered for some time. Ultimately, this will depend on the relative magnitude of the non-linearities, the range of modes that become unstable and the self-coupling. Substantial work has been performed analytically and numerically on plasma instabilities in QCD and QCD-like theories (see for instance \cite{sim1,sim2,sim3,sim4,sim5}), both in a HIC context and for simplified models. 

In the present work we will explore the instabilities of gauge field perturbations around homogeneous and constant E and B fields in pure classical SU(2) gauge theory. For simplicity, we will ignore any fermions coupled to the gauge fields, which may feel and enhance the anisotropy.  Our approach is inspired by \cite{bazak1,bazak2}, where the authors investigated the dispersion relations of perturbations in the linearized theory, in a variety of gauge field backgrounds. Part of the present work is an extension of their analysis, while later parts attempts to go beyond the linear approximation. 

\subsection{The equations of motion order by order}
\label{sec:eom}
The classical equation of motion for pure SU(2) gauge theory reads
\begin{eqnarray}
D^{ab}_\mu F^{\mu\nu,b}=j^{\nu,a},
\end{eqnarray}
where the covariant derivative is
\begin{eqnarray}
D_\mu^{ab}=\partial_\mu\delta^{ab}-g \epsilon^{abc}A_\mu^c,
\end{eqnarray}
and the field strength tensor is
\begin{eqnarray}
F^{\mu\nu,a}=\partial^\mu A^{\nu,a}-\partial^\nu A^{\mu,a}+g\epsilon^{abc}A^{\mu,b}A^{\nu,c}.
\end{eqnarray}
The gauge field $A_{\mu,a}$ is a 12-component vector, corresponding to colour indices $a=1,2,3$ and Lorentz indices $\mu=0,1,2,3$.  Because of gauge invariance, there is some redundancy in these degrees of freedom, and we will make use of this point shortly. The metric signature is taken to be $(+---)$, so that $x_0=x^0=t$ and $-x_i=x^i=(x,y,z)$, and $\partial^0=\partial_0=\partial/\partial t$, $-\partial^i=\partial_i=\partial/\partial x^i$. Then $E^{i,a}=F^{i0,a}$, while $B^{i,a}=\frac{1}{2}\epsilon^{ijk}F^{kj,a}$.

We will assume that the gauge field may be decomposed into a background field and a fluctuation, with the notation
\begin{eqnarray}
A^{\mu}_a(x,t)=\bar{A}^{\mu}_a(x,t)+h^{\mu}_a(x,t),
\end{eqnarray}
and with an assumption that the fluctuations are in some sense "small", $h_a^\mu\ll \bA^\mu_a$. It then makes sense to expand order by order in $h^\mu_a$.

\subsubsection*{Background field}
\label{sec:othorder}
The equation of motion for the background field (expanding to zero'th order in the fluctuation $h^\mu_a$) is simply
\begin{eqnarray}
\bar{D}^{ab}_\mu \bar{F}^{\mu\nu,b}=j^{\nu,a}
\label{eq:background},
\end{eqnarray}
where the bar on $\bar{D}^{ab}_\mu$ and $\bar{F}^{\mu\nu,b}$ indicate that they are expressed only in terms of the background field $\bA$.
We have assumed that the external current enters in this background field equation in its entirety. In practice, we will in the following state the background field $\bA^\mu_a$ and simply infer what the current $j^\nu_a$ is required to be to generate such a field. In principle, this current may be taken to arise from some distribution of colour-charged particles, but we will not discuss this further.

\subsubsection*{Linear in fluctuations}
\label{sec:1storder}
To linear order in fluctuations, one finds
\begin{eqnarray}
\label{eq:lineareom}
\partial_\mu\partial^\mu h^{\nu}_a+g\epsilon_{abc}\left(2\partial^\nu \bA_\mu^b h^\mu_c+2\partial_\mu \bA^\nu_c h^\mu_b+2 \bA^\mu_b\partial_\mu h^\nu_c\right)\nonumber\\
-g^2\left(2\bA_\mu^c\bA^\nu_a h^\mu_c-2\bA^\nu_c \bA_\mu^a h^\mu_c+\bA_\mu^c \bA^\mu_c h^\nu_a-\bA_\mu^c\bA^\mu_a h^\nu_c\right)=0.\nonumber\\
\end{eqnarray}
We adopt the Lorenz gauge $\partial_\mu\bar{A}^\mu_a=0$ for the background field, and have imposed the background gauge fixing condition on the fluctuations, $\bar{D}^{ab}_\mu h^{\mu,b}=0$, where $\bar{D}_\mu^{ab}$ again denotes the covariant derivative in terms of the background field $\bar{A}_{\mu}^b$ only. 
In a more compact notation, this may be rewritten as
\begin{eqnarray}
\left[g^{\mu\nu}(\bar{D}_\rho\bar{D}^\rho)_{ac}+2g \epsilon_{abc}\bar{F}_b^{\mu\nu}\right]h_{\nu,c}=\mathcal{M}^{\mu\nu}_{ac} [12\times 12] h_{\nu,c}=0.\nonumber\\
\end{eqnarray}
As we will see, in momentum space this turns into a $[12\times 12]$ matrix equation for $h^\mu_a$, which has unstable modes for certain choices of $\bA^\mu_a$. Some of these correspond to non-zero homogeneous colour-electric and/or -magnetic fields, and we will in section \ref{sec:background} consider a fairly broad subset. Given $\bA^\mu_a$, one may compute the dispersion relations for these modes.

\subsubsection*{Quadratic in fluctuations}
\label{sec:2ndorder}
At early times, when $h^\mu_a$ is truly $\ll \bA^\mu_a$, the linear approximation is valid and unstable modes grow exponentially. However, non-linearities will eventually impact the evolution. At second order in fluctuations, we have
\begin{eqnarray}
\label{eq:quadratic}
&&g\epsilon_{abc}\left(2\partial_\mu h^{\nu}_ch^{\mu}_b+\partial^\nu h_\mu^b h^{\mu}_c\right)\nonumber\\
&&-g^2\left(
2\bA^\mu_c h_\mu^c h^\nu_a
-2 \bA^\mu_a h_\mu^ch^\nu_c
+\bA^\nu_a h_\mu^ch^\mu_c
-\bA^\nu_ch_\mu^c h^\mu_a
\right).\nonumber\\
\end{eqnarray}
which upon taking the expectation values for the fluctuations, and truncating at quadratic order, naturally adds to the background equation of motion (\ref{eq:background}). Defining
\begin{eqnarray}
\label{eq:quadratic2}
C^{\mu\nu}_{ab}=\langle h^\mu_a h^\nu_b\rangle,\quad \bar{j}^{\nu}_{a}=g\epsilon_{abc}\langle 2\partial_\mu h^\nu_{c} h^\mu_b+\partial^\nu h_\mu^b  h^\mu_c\rangle,
\end{eqnarray}
we have schematically
\begin{eqnarray}
\bar{D}^{ab}_\mu \bar{F}^{\mu\nu,b}
-g^2\bar{M}^{\nu\mu}_{ab}[C]\bA_{\mu,b}=j^{\nu}_a - \bar{j}^\nu_a.
\label{eq:quadraticcont}
\end{eqnarray}
where $\bar{M}$ is some matrix involving the $C$ correlators in (\ref{eq:quadratic2}). This in effect changes both the background equation of motion and the effective current, and if $j^\nu_a$ is constant will eventually ruin the assumption of a constant background electric/magnetic fields. We will investigate the effect of this non-linear back reaction below.

\subsubsection*{Cubic in fluctuations}
\label{sec:3rdorder}
Finally, the equation of motion has cubic contributions in the fluctuations, which read
\begin{eqnarray}
-g^2\left[
h^{\nu}_ah_\mu^ch^{\mu}_c-h^{\mu}_ah_\mu^ch^{\nu}_c
\right],
\end{eqnarray}
and which again upon taking expectation values naturally add to the linear equation for the modes as
\begin{eqnarray}
\label{eq:cubiccont}
\left[g^{\mu\nu}(\bar{D}_\rho\bar{D}^\rho)_{ac}+2g \epsilon_{abc}\bar{F}_b^{\mu\nu}-g^2\hat{M}^{\mu\nu}_{ac}(C)\right]h_{\nu,c}
=0.\nonumber\\
\end{eqnarray}
The matrix $\hat{M}$ is expressed in terms if the $C$ correlators, providing a corrected linear equation for the fluctuations $h^{\mu}_a$. This gives rise to new time-dependent dispersion relations, altering the pattern of instabilities. 

\subsection{This work}
In section \ref{sec:background}, we will consider a fairly general set of homogeneous, anisotropic field backgrounds. For three of these we will in section \ref{sec:dispersion} compute the dispersion relations of the fluctuation modes explicitly, and establish for which momentum regions the fluctuations are unstable. The remaining cases are more involved, and we will simply sketch how one would go about finding dispersion relations for them in Appendix \ref{app:4to10}, relying heavily on \cite{bazak1,bazak2}. In section \ref{sec:nonlinear} we choose one of the three cases and compute the eigenmodes explicitly, in order to compute the $C$ correlators. This will in turn allow us to compute the cubic correction matrix $\hat{M}$ and the quadratic corrections $\bar{M}$ and $\bar{J}$, from which we will be able to study the evolution of the instabilities. We provide some discussion, conclusions and suggestions for future work in section \ref{sec:conclusion}.

\section{Anisotropic backgrounds with constant, homogeneous electric/magnetic fields}
\label{sec:background}

\subsection{Electric field only}
We will first consider a constant, homogeneous colour-electric field only, and set up our procedure and notation. Firstly, without loss of generality, we can choose the electric field in the $i=1$, $a=1$ direction. We hence require that
\begin{eqnarray}
\bar{F}^{i0}_a=E^{i}_a=\delta^{i1}_{a1}E.
\end{eqnarray}
Since 
\begin{eqnarray}
\bar{F}^{i0}_a = \partial^i\bA^0_a-\partial^0 \bA^i_a+ g\epsilon_{abc}\bA^i_b\bA^0_c,
\end{eqnarray}
such an electric field may correspond to a multitude of choices for $\bA^\mu_a$ (some of which may be gauge-equivalent). We choose to restrict ourselves to two cases: One where only the linear-in-$A$ ("Abelian") derivative part of $\bar{F}$ is non-zero. And one where only the quadratic-in-$A$ ("non-Abelian") part is non-zero. 
A homogeneous, time-independent {\it Abelian} electric field may then be written (in a compact notation we will use repeatedly for the 12-component objects such as $\bA^\mu_a$, $j^\mu_a$), as 
\begin{eqnarray}
\bA^\mu_a&&= \left[
\left(\begin{array}{c} \bA^0_1\\\bA^0_2\\\bA^0_3\end{array}\right)
\left(\begin{array}{c} \bA^1_1\\\bA^1_2\\\bA^1_3\end{array}\right)
\left(\begin{array}{c} \bA^2_1\\\bA^2_2\\\bA^2_3\end{array}\right)
\left(\begin{array}{c} \bA^3_1\\\bA^3_2\\\bA^3_3\end{array}\right)
\right]\nonumber\\
&&=
 \left[
\left(\begin{array}{c} -x\,C_1\\0\\0\end{array}\right)
\left(\begin{array}{c} -t\,C_2\\0\\0\end{array}\right)
\left(\begin{array}{c} 0\\0\\0\end{array}\right)
\left(\begin{array}{c} 0\\0\\0\end{array}\right)
\right],\nonumber\\
\end{eqnarray}
so that
\begin{eqnarray}
E^1_1\equiv E=C_1+C_2.
\end{eqnarray}
This corresponds to a vanishing current $j^{\nu,a}=0$. 
Similarly a homogeneous time-independent {\it non-Abelian} electric field may be written as
\begin{eqnarray}
\bA^\mu_a=
 \left[
\left(\begin{array}{c} 0\\D_2\\D_1\end{array}\right)
\left(\begin{array}{c} 0\\D_4\\D_3\end{array}\right)
\left(\begin{array}{c} 0\\0\\0\end{array}\right)
\left(\begin{array}{c} 0\\0\\0\end{array}\right)
\right],
\end{eqnarray}
to find
\begin{eqnarray}
E^1_1\equiv E=g(D_4D_1-D_3D_2).
\end{eqnarray}
This corresponds to a current with four non-vanishing components:
\begin{eqnarray}
j^{\nu}_a=-gE
 \left[
\left(\begin{array}{c} 0\\D_3\\-D_4\end{array}\right)
\left(\begin{array}{c} 0\\D_1\\-D_2\end{array}\right)
\left(\begin{array}{c} 0\\0\\0\end{array}\right)
\left(\begin{array}{c} 0\\0\\0\end{array}\right)
\right].\nonumber\\
\end{eqnarray}

\subsection{Magnetic field only}
For a homogeneous magnetic field we can again choose the $\mu=1$, $a=1$ direction, and write
\begin{eqnarray}
\frac{1}{2}\epsilon^{ijk}F^{kj}_a=B_a^i=\delta^{i1}_{a1}B.
\end{eqnarray}
An {\it Abelian} background field can be parametrized as 
\begin{eqnarray}
\bA^\mu_a=
\left[
\left(\begin{array}{c} 0\\0\\0\end{array}\right)
\left(\begin{array}{c} 0\\0\\0\end{array}\right)
\left(\begin{array}{c} -z\,C_4\\0\\0\end{array}\right)
\left(\begin{array}{c} y\,C_3\\0\\0\end{array}\right)
\right],
\end{eqnarray}
so that
\begin{eqnarray}
 B^1_1=B=C_3+C_4.
\end{eqnarray}
This again corresponds to a vanishing current $j^{\nu,a}=0$. The {\it non-Abelian} realisation is
\begin{eqnarray}
\bA^\mu_a=
\left[
\left(\begin{array}{c} 0\\0\\0\end{array}\right)
\left(\begin{array}{c} 0\\0\\0\end{array}\right)
\left(\begin{array}{c} 0\\D_6\\D_5\end{array}\right)
\left(\begin{array}{c} 0\\D_8\\D_7\end{array}\right)
\right],
\end{eqnarray}
with
\begin{eqnarray}
B^1_1= B=g(D_5D_8-D_6D_7).
\end{eqnarray}
This again corresponds to a current with four non-vanishing components:
\begin{eqnarray}
j^{\nu}_a=-gB
\left[
\left(\begin{array}{c} 0\\0\\0\end{array}\right)
\left(\begin{array}{c} 0\\0\\0\end{array}\right)
\left(\begin{array}{c} 0\\D_7\\-D_8\end{array}\right)
\left(\begin{array}{c} 0\\-D_5\\D_6\end{array}\right)
\right].
\end{eqnarray}
These expressions are minor generalisations to \cite{bazak1}, where the authors consider Abelian $E$ with $C_1=E$, $C_2=0$, Abelian $B$ with $C_3=B$, $C_4=0$, non-Abelian $E$ with $D_2=-D_3=\sqrt{E/g}$, $D_1=D_4=0$ and non-Abelian $B$ with $D_5=D_8=\sqrt{B/g}$, $D_6=D_7=0$. In \cite{bazak2} the same authors in addition considered arbitrary $D_2,D_3$, and $D_5, D_8$, respectively. We see, that there are a host of other possibilities for choosing the $C_i$ or the $D_i$ for a given $E$ or $B$. It will not be possible to keep full generality in the following, but we will further investigate some of the combinations in terms of the resulting dispersion relations. 

\subsection{Electric and magnetic fields, combined}

The natural generalisation is to consider the combination of an electric field $E^{1}_1$ together with a magnetic field that is/is not aligned in space and/or colour. Allowing for all combinations of Abelian/non-Abelian $\bar{A}^\mu_a$ one might expect 16 distinct cases. Some combinations are however not possible since additional components of magnetic and electric field are sourced. The task is therefore to identify vector potentials with the property that they generate only $E^1_1$ and then some combination of $B^i_a$, enforcing that they are either of Abelian or non-Abelian type, and that $E$ and $B$ are constant in space and time. One then finds a limited set of options, which we will briefly review in the following. 

\subsubsection*{Abelian $E$ and Abelian $B$}
Choosing a vector potential of the form
\begin{eqnarray}
\bA^\mu_a=
\left[
\begin{pmatrix}
-x\,C_1\\\\\\0\\0\\
\end{pmatrix}
\begin{pmatrix}
-t\,C_2-y\,C_{16}\\+z\,C_9\\\\0\\0\\
\end{pmatrix}
\begin{pmatrix}
 -z\,C_{4}\\+x\,C_{15}\\\\0\\0\\
\end{pmatrix}
\begin{pmatrix}
-x\,C_{10}\\+y\,C_{3}\\\\0\\0\\
\end{pmatrix}
\right],\nonumber\\
\end{eqnarray}
gives Abelian $E$- and $B$-fields aligned in colour, with 
\begin{eqnarray}
&&E^{1}_1=C_1+C_2,\nonumber\\
&&B^i_a=g\left[
\left(\begin{array}{c} C_3+C_4\\0\\0\end{array}\right)
\left(\begin{array}{c} C_9+C_{10}\\0\\0\end{array}\right)
\left(\begin{array}{c} C_{15}+C_{16}\\0\\0\end{array}\right)
\right].\nonumber\\
\end{eqnarray}
Explicitly setting $\bA^0_1=0$ ($C_1=0$), this opens up a few more options:
\begin{eqnarray}
\bA^\mu_a=
\left[
\left(\begin{array}{c} 0\\0\\0\end{array}\right)
\left(\begin{array}{c} -t\,C_2-y\,C_{16}+z\,C_9\\-y\,C_{18}+z\,C_{11}\\-y\,C_{20}+z\,C_{13}\end{array}\right)
\left(\begin{array}{c} 0\\0\\0\end{array}\right)
\left(\begin{array}{c} 0\\0\\0\end{array}\right)
\right],\nonumber\\
\end{eqnarray}
which gives Abelian $E$- and $B$-fields orthogonal in space:
\begin{eqnarray}
&&E^{1}_1=C_2,\nonumber\\
&&B^i_a=g\left[
\left(\begin{array}{c} 0\\0\\0\end{array}\right)
\left(\begin{array}{c} C_9\\C_{11}\\C_{13}\end{array}\right)
\left(\begin{array}{c} C_{16}\\C_{18}\\C_{20}\end{array}\right)
\right].\nonumber\\
\end{eqnarray}

\subsubsection*{Abelian E and non-Abelian B}
A vector potential of the form
\begin{eqnarray}
\bA^\mu_a=
\left[
\left(\begin{array}{c} 0\\0\\0\end{array}\right)
\left(\begin{array}{c} -tC_2\\D_4\\D_3\end{array}\right)
\left(\begin{array}{c} D_{11}\\0\\0\end{array}\right)
\left(\begin{array}{c} D_{12}\\0\\0\end{array}\right)
\right],
\end{eqnarray}
gives Abelian $E$- and non-Abelian $B$-fields orthogonal in space and colour:
\begin{eqnarray}
&&E^{1}_1=C_2,\nonumber\\
&&B^i_a=g\left[
\left(\begin{array}{c} 0\\0\\0\end{array}\right)
\left(\begin{array}{c} 0\\D_3D_{12}\\-D_4D_{12}\end{array}\right)
\left(\begin{array}{c} 0\\-D_3D_{11}\\D_4D_{11}\end{array}\right)
\right].\nonumber\\
\end{eqnarray}

\subsubsection*{Non-Abelian E and non-Abelian B}
Finally, we may choose
\begin{eqnarray}
\bA^\mu_a=
\left[
\left(\begin{array}{c} 0\\0\\D_1\end{array}\right)
\left(\begin{array}{c} 0\\D_4\\0\end{array}\right)
\left(\begin{array}{c} 0\\0\\D_5\end{array}\right)
\left(\begin{array}{c} 0\\0\\D_7\end{array}\right)
\right],
\end{eqnarray}
which gives non-Abelian $E$- and $B$-fields orthogonal in colour, but aligned in space.
\begin{eqnarray}
&&E^{1}_1=gD_1D_4,\nonumber\\
&&B^i_a=g\left[
\left(\begin{array}{c} 0\\0\\0\end{array}\right)
\left(\begin{array}{c} D_4D_7\\0\\0\end{array}\right)
\left(\begin{array}{c} -D_4D_5\\0\\0\end{array}\right)
\right].\nonumber\\
\end{eqnarray}
The combination
\begin{eqnarray}
\bA^\mu_a=
\left[
\left(\begin{array}{c} 0\\D_2\\0\end{array}\right)
\left(\begin{array}{c} 0\\0\\D_3\end{array}\right)
\left(\begin{array}{c} 0\\D_6\\0\end{array}\right)
\left(\begin{array}{c} 0\\D_8\\0\end{array}\right)
\right],
\end{eqnarray}
is equivalent.

In summary, we have seen that we can find: 
\begin{itemize}
\item Abelian $E$ or $B$, or non-Abelian $E$ or $B$ with a somewhat generalised parametrisation of what was reported in \cite{bazak1}.
\item Space-aligned, colour-aligned $E$ and $B$: Possible as an Abelian/Abelian combination. $C_1\neq 0$, $C_3\neq 0$ is the special case reported in \cite{bazak1}. 
\item Space-aligned, colour-orthogonal $E$ and $B$: Can not be realised in our setup.
\item Space-orthogonal, colour-aligned $E$ and $B$: Possible as Abelian/Abelian and non-Abelian/non-Abelian combinations. 
\item Space-orthogonal, colour-orthogonal $E$ and $B$: Possible as Abelian/Abelian and Abelian/non-Abelian combinations.
\end{itemize}
We also see that there are several computationally distinct cases, depending on whether the vector potential is itself space- and/or time-dependent. We will press on with the three purely non-Abelian cases (E only, B only, E and B), where $\bA^\mu_a$ is homogeneous and constant. We refer to Appendix \ref{app:4to10} for a brief discussion of the issues involved for the Abelian vector potentials. 

\subsection{Non-Abelian B, E=0}
\label{sec:NAB}

Let us return to the case of a non-Abelian B-field and zero E-field.
 In 4-dimensional momentum space, the linear equation for the modes of the fluctuations reduce to a set of 12 coupled linear algebraic equations for $h^\mu_a$, which however split into 3 independent sections. 
\begin{eqnarray}
\label{eq:12by12}
\mathcal{M}^{\mu\nu}_{ac}\left[12\times 12\right]h_\nu^c=\left(\begin{array}{cccc}
M_{00}&0&0&0\\
0&M_{11}&0&0\\
0&0&M_{22}&M_{23}\\
0&0&M_{32}&M_{33}\\
\end{array}\right)
\left(\begin{array}{c}
h^0_{1,2,3}\\
h^1_{1,2,3}\\
h^2_{1,2,3}\\
h^3_{1,2,3}\\
\end{array}\right)
,\nonumber\\
\end{eqnarray}
where in addition $M_{00}=M_{11}=M_{22}=M_{33}$ are all hermitian, and $M_{23} =M_{32}^\dagger$, so that the whole matrix $M$ is hermitian.  
Hence the equations for $h^0_{1,2,3}$ ($M_{00}$) and $h^1_{1,2,3}$ ($M_{11}$) are identical 
\begin{widetext}
\begin{eqnarray}
\left(\begin{array}{ccc}
\Box_k+D_5^2+D_6^2+D_7^2+D_8^2&2i(D_5k_y+D_7k_z)&-2i(D_6k_y+D_8k_z)\\
-2i(D_5k_y+D_7 k_z)&\Box_k+D_5^2+D_7^2&-(D_5D_6+D_7D_8)\\
2i(D_6 k_y+D_8k_z)&-(D_5D_6+D_7D_8)&\Box_k+D_6^2+D_8^2
\end{array}\right)
\left(\begin{array}{c}
h^{0,1}_1\\
h^{0,1}_2\\
h^{0,1}_3\\
\end{array}\right)=0,\nonumber\\
\end{eqnarray}
while the $h^{2,3}_{1,2,3}$-sections mix ($M_{22,23,32,33}$)
\begin{eqnarray}
\left(\begin{array}{ccc}
\Box_k+D_5^2+D_6^2+D_7^2+D_8^2&2i(D_5k_y+D_7k_z)&-2i(D_6k_y+D_8k_z)\\
-2i(D_5k_y+D_7 k_z)&\Box_k+D_5^2+D_7^2&-(D_5D_6+D_7D_8)\\
2i(D_6 k_y+D_8k_z)&-(D_5D_6+D_7D_8)&\Box_k+D_6^2+D_8^2
\end{array}\right)
\left(\begin{array}{c}
h^{2}_1\\
h^{2}_2\\
h^{2}_3\\
\end{array}\right)
+
\left(\begin{array}{ccc}
0&0&0\\
0&0&-2B\\
0&2B&0
\end{array}\right)
\left(\begin{array}{c}
h^{3}_1\\
h^{3}_2\\
h^{3}_3\\
\end{array}\right)=0,\nonumber\\
\end{eqnarray}
and
\begin{eqnarray}
\left(\begin{array}{ccc}
\Box_k+D_5^2+D_6^2+D_7^2+D_8^2&2i(D_5k_y+D_7k_z)&-2i(D_6k_y+D_8k_z)\\
-2i(D_5k_y+D_7 k_z)&\Box_k+D_5^2+D_7^2&-(D_5D_6+D_7D_8)\\
2i(D_6 k_y+D_8k_z)&-(D_5D_6+D_7D_8)&\Box_k+D_6^2+D_8^2
\end{array}\right)
\left(\begin{array}{c}
h^{3}_1\\
h^{3}_2\\
h^{3}_3\\
\end{array}\right)
+
\left(\begin{array}{ccc}
0&0&0\\
0&0&2B\\
0&-2B&0
\end{array}\right)
\left(\begin{array}{c}
h^{2}_1\\
h^{2}_2\\
h^{2}_3\\
\end{array}\right)=0.\nonumber\\
\end{eqnarray}
\end{widetext}
To somewhat mitigate notational clutter we have absorbed $g$ into $\bar{A}$, $gD_i\rightarrow D_i$, and defined the box operator in momentum space $\Box_k=-\omega^2+k_x^2+k^2_T$, with $k_T^2=k_y^2+k_z^2$. We note that the $D_{5,6,7,8}$ do not only appear in the combination $B=D_5D_8-D_6D_7$, but depend on the $D_i$ individually.

\begin{figure}
\begin{center}
\includegraphics[width=0.35\textwidth]{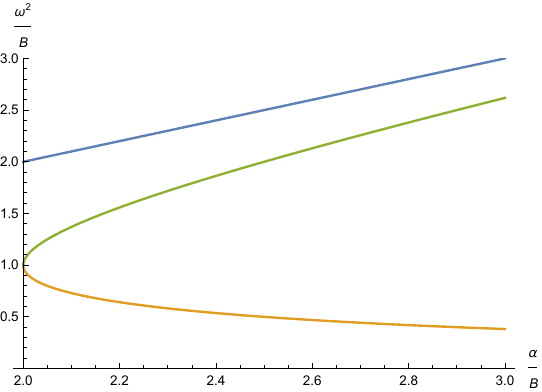}
\includegraphics[width=0.35\textwidth]{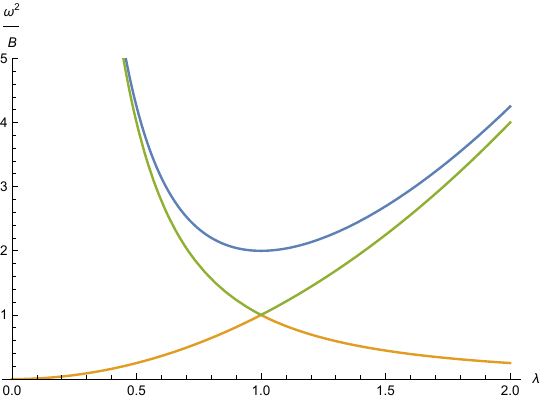}
\end{center}
\caption{Non-Abelian B: The three eigenvalues in the 0, 1 sectors for ${\bf k}=0$, general $D_{i}$ (top) and for $\sqrt{B}=D_6/\lambda=-D_7\lambda$ (bottom).}
\label{fig:NAB_k0}
\end{figure}
\subsubsection*{Non-Abelian B: 0 and 1 sector}

\begin{figure}
\begin{center}
\includegraphics[width=0.35\textwidth]{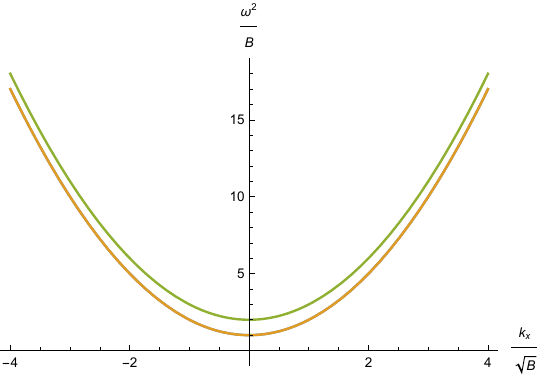}
\includegraphics[width=0.35\textwidth]{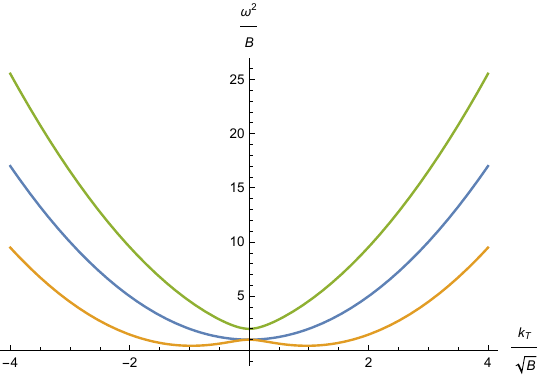}
\end{center}
\caption{Non-Abelian B: The three eigenvalues in the 0, 1 sectors for $\sqrt{B}=D_6=-D_7$, with $k_T=0$ (top) and $k_x=0$ (bottom).}
\label{fig:NAB_k_67}
\end{figure}
\begin{figure}
\begin{center}
\includegraphics[width=0.35\textwidth]{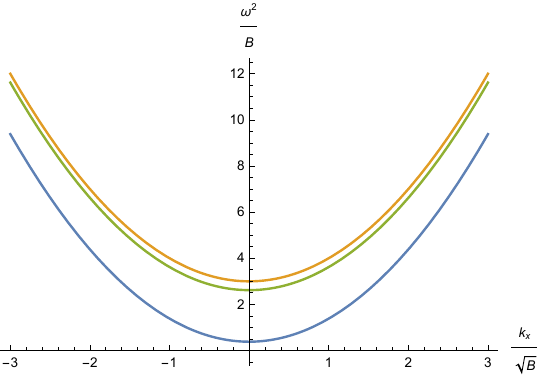}
\includegraphics[width=0.35\textwidth]{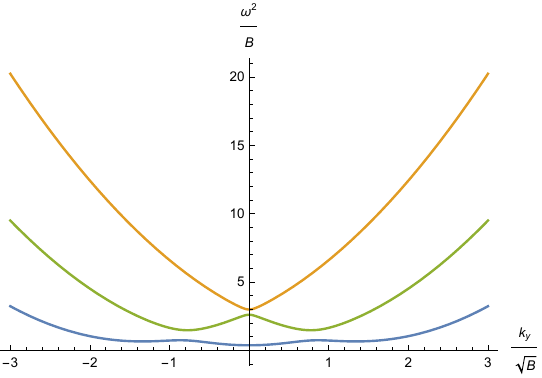}
\includegraphics[width=0.35\textwidth]{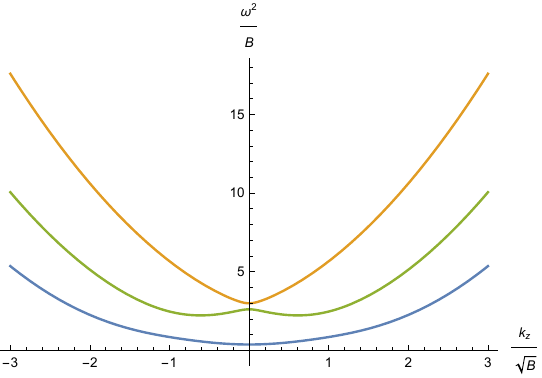}
\end{center}
\caption{Non-Abelian B: The three eigenvalues in the 0, 1 sectors for $\sqrt{B}=D_6=-D_7$, with non-zero $D_5=\sqrt{nB}$, $n=1$. For non-zero $k_x$ (top), $k_y$ (middle) and $k_z$ (bottom). }
\label{fig:NAB_k_567}
\end{figure}
The condition on $\omega$ to satisfy the equation is that $\textrm{Det}[M_{00}]=0=\textrm{Det}[M_{11}]$, which reduces to the cubic equation
\begin{eqnarray}
\Box_k^3+2\alpha_B\Box_k^2+\beta_B\Box_k+\gamma_B=0,
\label{eq:eq01}
\end{eqnarray}
with
\begin{eqnarray}
\alpha_B &=& (D_5^2+D_6^2+D_7^2+D_8^2)\geq 2B,\\
\beta_B&=& \alpha_B^2+B^2-4\left(
(D_5k_y+D_7k_z)^2+(D_6k_y+D_8k_z)^2
\right),\nonumber\\
\\
\gamma_B&=&B^2(\alpha_B-4k_T^2).
\end{eqnarray}
Let us first consider the zero-momentum case $k_T=k_x=0$, so that $\omega^2=-\Box_k$, and 
\begin{eqnarray}
-\left(\omega^2\right)^3+2\alpha_B\left(\omega^2\right)^2-\left(\alpha_B^2+B^2\right)\omega^2+B^2\alpha_B=0.\nonumber\\
\end{eqnarray}
The solutions for $\omega^2$ only depend on the combination $\alpha_B/B$,
\begin{eqnarray}
\frac{\omega^2}{B}=\frac{\alpha_B}{B},\quad \frac{\omega^2}{B}=\frac{1}{2}\left(\frac{\alpha_B}{B}\pm \sqrt{\frac{\alpha_B^2}{B^2}-4}\right),
\end{eqnarray}
and they are all real and positive (see Figure \ref{fig:NAB_k0}, top). For instance in the case  $D_5=D_8=0$, $D_6=-D_7=\sqrt{B}$, we find $\alpha_B/B=2$ so that the eigenvalues reduce to $\omega^2=(2B,B,B)$ . For the more general case $D_6=\sqrt{B}\lambda$, $-D_7=\sqrt{B}/\lambda$ for some $\lambda$ \cite{bazak1}, $\alpha_B/B=\lambda^2+\lambda^{-2}\geq 2$, and the solutions are again real and positive (see Figure \ref{fig:NAB_k0}, bottom). As a sanity check, we note that when $B=0$, the eigenvalues are $\omega^2=(0,\alpha_B,\alpha_B)$. These are real and positive, and only zero when all the $D_i$ are zero.

\begin{figure}
\begin{center}
\includegraphics[width=0.35\textwidth]{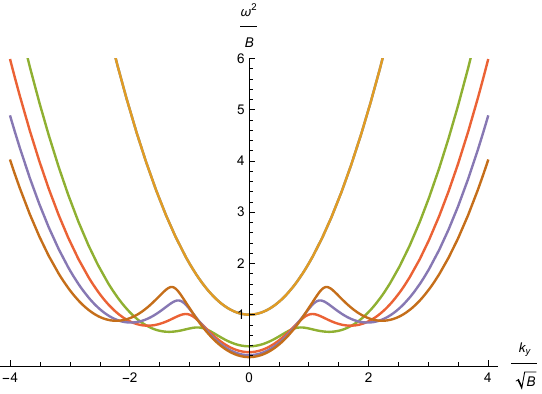}
\includegraphics[width=0.35\textwidth]{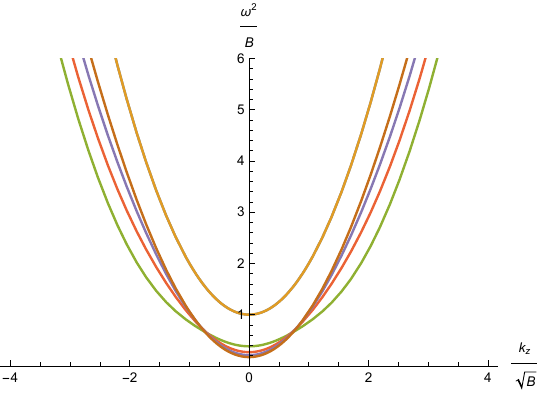}
\end{center}
\caption{Non-Abelian B: The lowest eigenvalues in the 0, 1 sectors for $\sqrt{B}=D_6=-D_7$, with non-zero $D_5=\sqrt{nB}$, $n=0,1,2,3,4$. For non-zero $k_y$ (top) and $k_z$ (bottom). }
\label{fig:NAB_k_567_n}
\end{figure}
For non-zero momentum $k_T,k_x$, the explicit realisation of $D_{5,..,8}$ becomes more important. Again for the case $D_6=-D_7=\sqrt{B}$ one finds ($\alpha_B/B=2$, $\beta_B/B^2=5-4k_T^2/B$, $\gamma_B/B^3=2-4k_T^2/B$),
\begin{eqnarray}
\label{eq:Bspecial1}
&&\left(-\frac{\omega^2}{B}+\frac{k_T^2}{B}+\frac{k_x^2}{B}\right)^3+4\left(-\frac{\omega^2}{B}+\frac{k_T^2}{B}+\frac{k_x^2}{B}\right)^2\nonumber\\&&+\left(5-4\frac{k_T^2}{B}\right)\left(-\frac{\omega^2}{B}+\frac{k_T^2}{B}+\frac{k_x^2}{B}\right)+2-4\frac{k_T^2}{B}=0,\nonumber\\
\end{eqnarray}
which has the solutions (see Figure \ref{fig:NAB_k_67})
\begin{eqnarray}
\frac{\omega^2}{B}=1+\frac{k_x^2}{B}+\frac{k_T^2}{B},\quad \frac{\omega^2}{B} =\frac{3}{2} +\frac{k_x^2}{B}+\frac{k_T^2}{B}\pm\frac{1}{2}\sqrt{1+16\frac{k_T^2}{B}}.\nonumber\\
\end{eqnarray}
These are real and positive for all $B,k_x,k_T$, and hence the 0 and 1 components of the fluctuations are stable for this choice of $D_{5,..,8}$.

For fixed $B$, a 3-dimensional space of realisations of $D_{5,..,8}$ is available, which we will not fully explore here. We will however briefly consider again the case $D_6=-D_7=\sqrt{B}$, fixing $D_8=0$, allowing us to choose any value for $D_5$ without changing $B$. We choose $D_5=\sqrt{nB}$, and vary $n$. In Figure \ref{fig:NAB_k_567} we show the eigenvalues for $n=1$, and their dependence on $k_{x,y,z}$ (respectively, keeping the other two constant). We note how the $k_T=\sqrt{k_y^2+k_z^2}$-dependence in Figure \ref{fig:NAB_k_67} is resolved into different dependence on $k_y$ and $k_z$. We can further examine the $n$-dependence of the lowest eigenvalue, as function of $k_y$ and $k_z$, shown in Figure \ref{fig:NAB_k_567_n}. We see a non-trivial dependence, but the eigenvalue remains real and positive. 

\subsubsection*{Non-Abelian B: 2 and 3 sector}

\begin{figure}
\begin{center}
\includegraphics[width=0.35\textwidth]{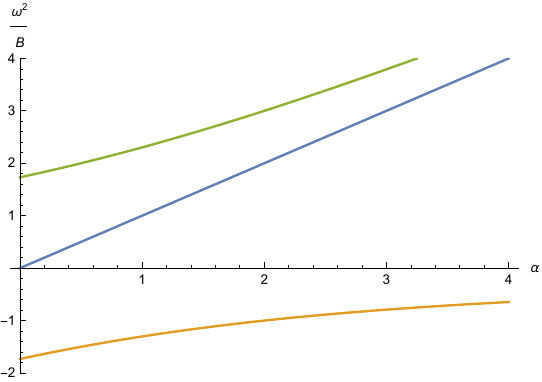}
\includegraphics[width=0.35\textwidth]{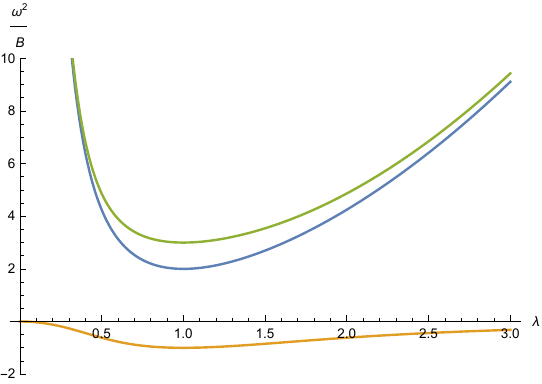}
\end{center}
\caption{Non-Abelian B: The three double eigenvalues in the 2, 3 sectors for ${\bf k}=0$, general $D_{i}$ (top) and for $\sqrt{B}=D_6/\lambda=-D_7\lambda$ (bottom).}
\label{fig:NAB23_k0}
\end{figure}
\begin{figure}
\begin{center}
\includegraphics[width=0.35\textwidth]{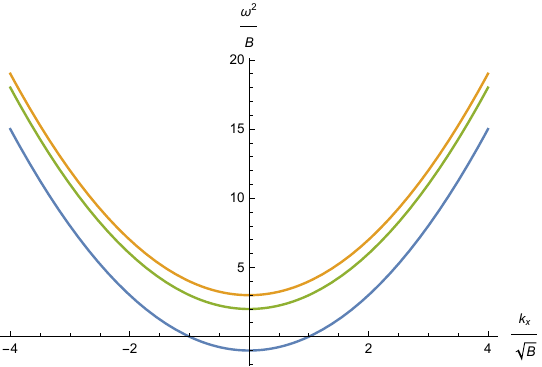}
\includegraphics[width=0.35\textwidth]{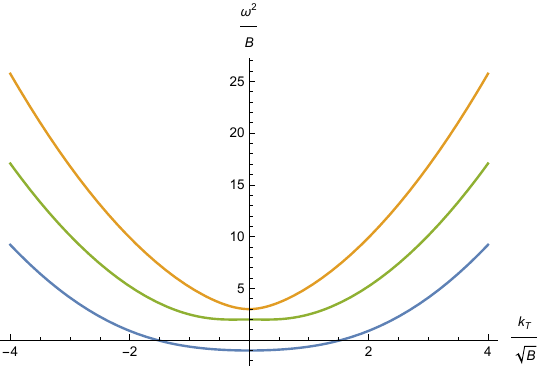}
\end{center}
\caption{Non-Abelian B: The three eigenvalues in the 2, 3 sectors for $\sqrt{B}=D_6=-D_7$, with $k_T=0$ (top) and $k_x=0$ (bottom).}
\label{fig:NAB_k_67_23}
\end{figure}
For the 2-3 component, the requirement $\textrm{Det}[M_{22, 23, 32, 33}]=0$ amounts to (compare to (\ref{eq:eq01}))
\begin{eqnarray}
\left[
\Box_k^3+2\alpha_B\Box_k^2+(\beta_B-4B^2)\Box_k+\gamma_B-4B^2\alpha_B.
\right]^2=0,\nonumber\\
\label{eq:B2233}
\end{eqnarray}
We may focus on the expression inside the square bracket, and infer that any solution is a double solution. 
For $k_T,k_x=0$ this becomes
\begin{eqnarray}
-(\omega^2)^3+2\alpha_B(\omega^2)^2-(\alpha_B^2-3B^2)\omega^2-3B^2\alpha_B=0,\nonumber\\
\end{eqnarray}
which has solutions
\begin{eqnarray}
\frac{\omega^2}{B}=\frac{\alpha_B}{B},\quad \omega^2=\frac{1}{2}\left(\frac{\alpha_B}{B}\pm \sqrt{\frac{\alpha_B^2}{B^2}+12}\right).
\label{eq:B2233om}
\end{eqnarray}
One of these eigenvalues is negative for all $\alpha_B/B$ (see Figure \ref{fig:NAB23_k0}, top). In particular for $D_5=D_8=0$, $D_6=-D_7=\sqrt{B}$ we find $\omega^2=(3B,2B,-B)$, while the parametrization $D_6=\sqrt{B}\lambda$, $-D_7=\sqrt{B}/\lambda$ gives the negative eigenvalue for $\omega^2$ in the interval $[-1,0]$ (see Figure \ref{fig:NAB23_k0}, bottom). Again, when $B=0$, $\omega^2=(0, \alpha_B, \alpha_B)$.

\begin{figure}
\begin{center}
\includegraphics[width=0.35\textwidth]{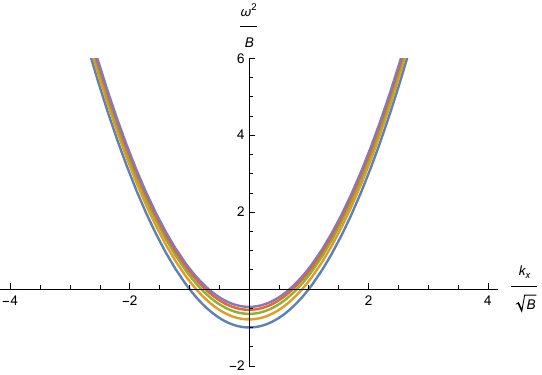}
\includegraphics[width=0.35\textwidth]{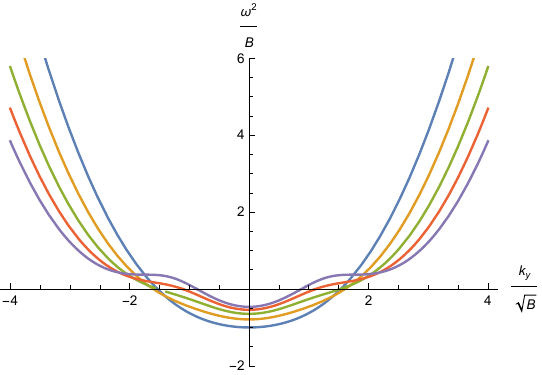}
\includegraphics[width=0.35\textwidth]{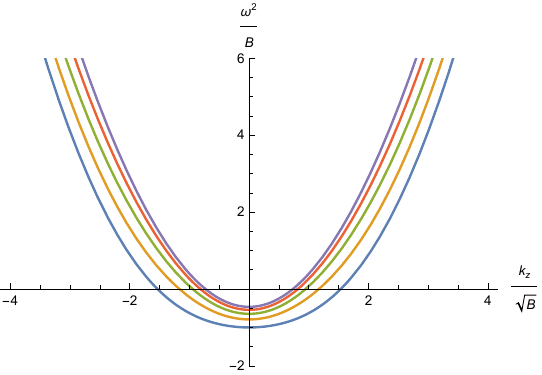}
\end{center}
\caption{Non-Abelian B: The lowest (negative) eigenvalue in the 2, 3 sectors for $\sqrt{B}=D_6=-D_7$, with non-zero $D_5=\sqrt{nB}$, $n=0,1,2,3,4$. For non-zero $k_x$ (top) $k_y$ (middle) and $k_z$ (bottom). }
\label{fig:NAB_k_567_n_23}
\end{figure}
For non-zero momentum, we consider again $D_6=-D_7=\sqrt{B}$, for which we need to solve
\begin{eqnarray}
\label{eq:Bspecial2}
&&\left(-\frac{\omega^2}{B}+\frac{k_T^2}{B}+\frac{k_x^2}{B}\right)^3+4\left(-\frac{\omega^2}{B}+\frac{k_T^2}{B}+\frac{k_x^2}{B}\right)^2\nonumber\\&&+\left(1-4\frac{k_T^2}{B}\right)\left(-\frac{\omega^2}{B}+\frac{k_T^2}{B}+\frac{k_x^2}{B}\right)-6-4\frac{k_T^2}{B}=0,\nonumber\\
\end{eqnarray}
This is a general cubic equation, which may be solved straightforwardly, to reveal (Figure \ref{fig:NAB_k_67_23}) that the (double) unstable mode becomes stable as $k_x$ and/or $k_T$ increase. Explicitly, for $k_T=0$, the unstable region is $|k_x|<1$, while for $k_x=0$ it is $|k_T|<\sqrt{(3-\sqrt{8})^{1/3}+(3+\sqrt{8})^{1/3}}\simeq 1.53$.

We can again extend our analysis somewhat, by allowing $D_5=\sqrt{n}B$, while keeping $D_8=0$. In Figure \ref{fig:NAB_k_567_n_23} we show the negative eigenvalue and its momentum dependence (nonzero $k_x$, $k_y$ or $k_z$, respectively), for different values of $n=0,1,2,3,4$. We see that larger $n$ reduces the region of instability, and we find that for $n\rightarrow \infty$, $\omega^2({\bf k=0})\rightarrow 0$.

Hence, in the presence of a constant, homogeneous non-Abelian magnetic field, for this choice of $D_{5,..,8}$, of the 12 sets of momentum modes present, two are potentially unstable, growing as
\begin{eqnarray}
\propto \exp{\left(\pm |\omega|t\right)},
\end{eqnarray}
with an $\omega$ given by the two lowest, degenerate solutions to (\ref{eq:B2233}) (exemplified by the $-$ solution in the special case of (\ref{eq:B2233om})). Furthermore, the region of instabilibity is centered around the origin (${\bf k}=0$), and that for large enough ${\bf k}$, these modes are again stable. Hence only a finite IR region of momentum modes are unstable, and this region is not only dependent on $B$, but also on the concrete choice of $D_{5,..,8}$, exemplified here by varying $D_5$.

In section \ref{sec:nonlinear} we will further find the eigenmodes of the system for non-Abelian B only, with $\sqrt{B}=D_6=-D_7$, $D_5=D_8=0$, compute the quadratic and cubic corrections to the evolution equations and solve them including these corrections.

\subsection{Non-Abelian E, B=0}
\label{sec:NAE}

For a non-Abelian background E-field and zero B-field, the linear equation for the Fourier modes again splits into 3 independent sections. The section for $h^2_{1,2,3}$ ($M_2$) and $h^3_{1,2,3}$ ($M_3$) are now identical and decoupled from each other, and read (absorbing again $g$ as $gD_i\rightarrow D_i$)
\begin{widetext}
\begin{eqnarray}
\left(\begin{array}{ccc}
\Box_k-D_1^2-D_2^2+D_3^2+D_4^2&-2i(D_1\omega-D_3k_x)&2i(D_2\omega-D_4 k_x)\\
+2i(D_1\omega-D_3k_x)&\Box_k-D_1^2+D_3^2&D_1D_2-D_3D_4\\
-2i(D_2\omega-D_4 k_x)&D_1D_2-D_3D_4&\Box_k-D_2^2+D_4^2
\end{array}\right)
\left(\begin{array}{c}
h^{2,3}_1\\
h^{2,3}_2\\
h^{2,3}_3\\
\end{array}\right)=0.\nonumber\\
\end{eqnarray}
while the $h^{0,1}_{1,2,3}$-sections mix ($M_{00},M_{01},M_{10},M_{11}$)
\begin{eqnarray}
\left(\begin{array}{ccc}
\Box_k-D_1^2-D_2^2+D_3^2+D_4^2&-2i(D_1\omega-D_3k_x)&2i(D_2\omega-D_4 k_x)\\
2i(D_1\omega-D_3k_x)&\Box_k-D_1^2+D_3^2&D_1D_2-D_3D_4\\
-2i(D_2\omega-D_4 k_x)&D_1D_2-D_3D_4&\Box_k-D_2^2+D_4^2
\end{array}\right)
\left(\begin{array}{c}
h^{0}_1\\
h^{0}_2\\
h^{0}_3\\
\end{array}\right)+
\left(\begin{array}{ccc}
0&0&0\\
0&0&-2E\\
0&2E&0
\end{array}\right)
\left(\begin{array}{c}
h^{1}_1\\
h^{1}_2\\
h^{1}_3\\
\end{array}\right)=0,\nonumber\\
\end{eqnarray}
and
\begin{eqnarray}
\left(\begin{array}{ccc}
\Box_k-D_1^2-D_2^2+D_3^2+D_4^2&-2i(D_1\omega-D_3k_x)&2i(D_2\omega-D_4 k_x)\\
2i(D_1\omega-D_3k_x)&\Box_k-D_1^2+D_3^2&D_1D_2-D_3D_4\\
-2i(D_2\omega-D_4 k_x)&D_1D_2-D_3D_4&\Box_k-D_2^2+D_4^2
\end{array}\right)
\left(\begin{array}{c}
h^{1}_1\\
h^{1}_2\\
h^{1}_3\\
\end{array}\right)+
\left(\begin{array}{ccc}
0&0&0\\
0&0&-2E\\
0&2E&0
\end{array}\right)
\left(\begin{array}{c}
h^{0}_1\\
h^{0}_2\\
h^{0}_3\\
\end{array}\right)=0.\nonumber\\
\end{eqnarray}
\end{widetext}

\subsubsection*{Non-Abelian E: 2 and 3 sector}

\begin{figure}
\begin{center}
\includegraphics[width=0.35\textwidth]{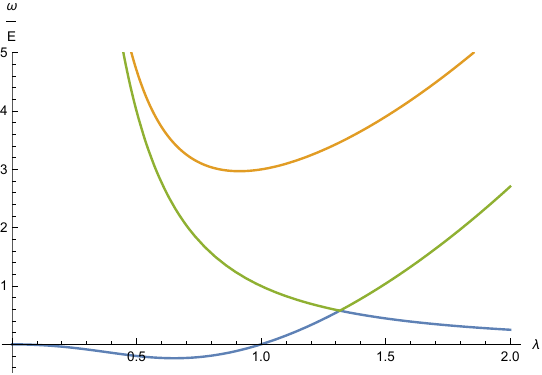}
\end{center}
\caption{Non-Abelian E: The three eigenvalues in the 2, 3 sectors for $\sqrt{E}=D_1/\lambda=D_4\lambda$.}
\label{fig:NAE23_k0_1}
\end{figure}
\begin{figure}
\begin{center}
\includegraphics[width=0.45\textwidth]{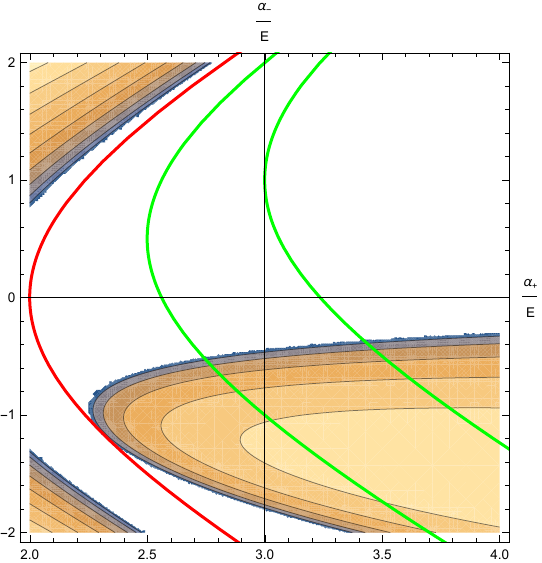}
\end{center}
\caption{Non-Abelian E: Regions of non-zero imaginary part of the eigenvalues in $\alpha_+/\alpha_-$ space. Overlaid the curve $\sqrt{E}=D_1/\lambda=D_4\lambda$ (red) and curves when further adding non-zero $D_3=\sqrt{nB}$, $n=0.5$, $1$ (green).}
\label{fig:NAE23_k0_2}
\end{figure}
The condition on $\omega$ is now that $\textrm{Det}[M_{22}]=0=\textrm{Det}[M_{33}]$, which is again a cubic equation of the form
\begin{eqnarray}
\Box_k^3+2\alpha_-\Box_k^2+\beta_E\Box_k+\gamma_E=0,
\label{eq:E2233}
\end{eqnarray}
where still $\Box_k=-\omega^2+k_x^2+k_T^2$, $k_T^2=k_y^2+k_z^2$, and
\begin{eqnarray}
\label{eq:NAEabg}
\alpha_-&=&-D_1^2-D_2^2+D_3^2+D_4^2,\\
\alpha_+&=&D_1^2+D_2^2+D_3^2+D_4^2,\\
\beta_E&=&\alpha_-^2-E^2-4\left((D_1\omega-D_3 k_x)^2+(D_2\omega-D_4k_x)^2\right),\nonumber\\\\
\gamma_E&=&-E^2(\alpha_--4(k_x^2-\omega^2)),
\end{eqnarray}
keeping in mind that $E=D_1D_4-D_2D_3$. For $k_T=k_x=0$, we obtain\footnote{Note that because $\omega$ appears not only in the $\Box_k$, this is different from the $B\neq 0$, $E=0$ case.}
\begin{eqnarray}
-(\omega^2)^3+2\alpha_+(\omega^2)^2-(\alpha_-^2+3E^2)\omega^2-E^2\alpha_-=0.\nonumber\\
\end{eqnarray} 
There is now one real solution for $\omega^2$, while the other two solutions are in general non-real. Hence already in the $2$ and $3$ sectors, there are unstable modes in the presence of finite $E$. However, for the special case $D_1=D_4=\sqrt{E}$, $D_2=D_3=0$, the eigenvalues are $\omega^2=(0,E,3E)$, while for the generalisation $D_1=\sqrt{E}\lambda$, $D_4=\sqrt{E}/\lambda$, one of the eigenvalues is negative in the interval $\lambda\in [0,1]$ and the other two are real and positive  (see Figure \ref{fig:NAE23_k0_1}). It appears that the curve traced out by $D_1D_4=E$ in the space of $D_{1,..,4}$ is quite special, avoiding the regions where the eigenvalues are non-real. In Figure \ref{fig:NAE23_k0_2}, we show the $\alpha_+/\alpha_-$ plane, and the contours of non-zero imaginary value of the eigenvalues. Overlaid in red, the curve traced out at $\lambda$ is varied, which precisely avoids those regions. Overlaid in green, the curves traced out by adding non-zero $D_3=\sqrt{nE}$, $n=0.5, 1$. As another sanity check, we note that for $E=0$, $\omega^2=(0,\alpha_+\pm\sqrt{\alpha_+^2-\alpha_-^2})$, which is real and positive, but non-zero whenever the $D_i$ are.

\begin{figure}
\begin{center}
\includegraphics[width=0.35\textwidth]{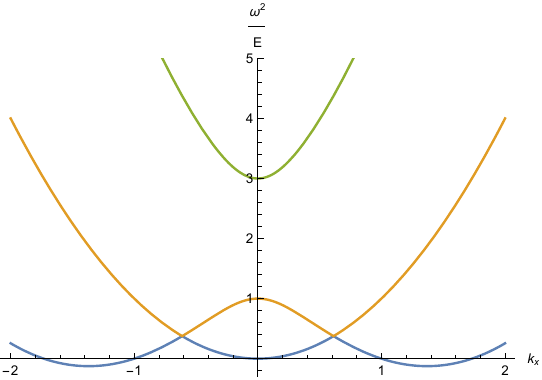}
\includegraphics[width=0.35\textwidth]{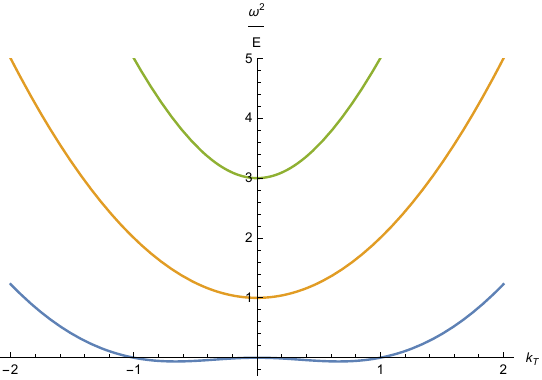}
\includegraphics[width=0.45\textwidth]{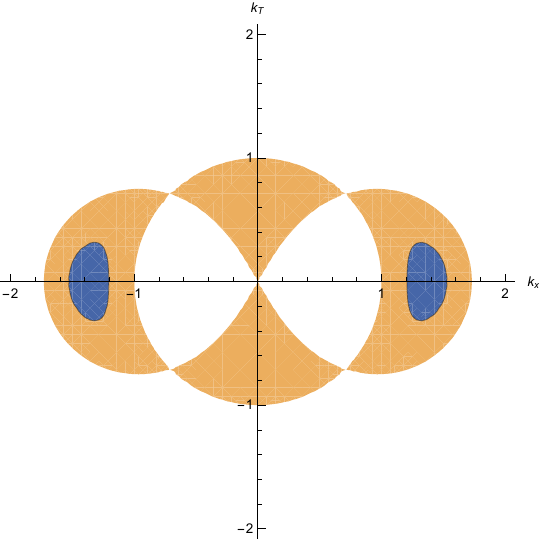}
\end{center}
\caption{Non-Abelian E: The three eigenvalues in the 2, 3 sectors for $\sqrt{E}=D_1=D_4$, as a function $k_x$ (top) and $k_T$ (middle), and the region in the $k_x/k_T$-plane, where the lowest eigenvalue is negative (bottom).}
\label{fig:NAE23_Dspec_k}
\end{figure}
For non-zero momentum, we immediately specialize to $D_1=D_4=\sqrt{E}$, $D_2=D_3=0$, so that we need to solve
\begin{eqnarray}
&&\left(-\frac{\omega^2}{E}+\frac{k_T^2}{E}+\frac{k_x^2}{E}\right)\times\nonumber\\&&
\qquad\left[
\left(-\frac{\omega^2}{E}+\frac{k_T^2}{E}+\frac{k_x^2}{E}\right)^2-4\left(\frac{\omega^2}{E}+\frac{k_x^2}{E}\right)-1\right]
\nonumber\\&&
\qquad\qquad\qquad-4\left(\frac{\omega^2}{E}-\frac{k_x^2}{E}\right)=0.\nonumber\\
\end{eqnarray}
The solutions are real, and one of them is negative for some regions of $(k_x,k_T^2)$, as shown in Figure \ref{fig:NAE23_Dspec_k}. 

\subsubsection*{Non-Abelian E: 0 and 1 sector}

\begin{figure}
\begin{center}
\includegraphics[width=0.45\textwidth]{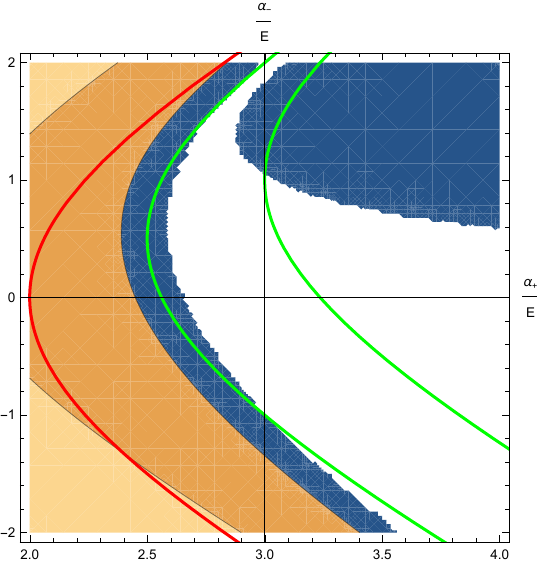}
\end{center}
\caption{Non-Abelian E: Regions of non-zero imaginary part of the eigenvalues in the 0, 1 sector in $\alpha_+/\alpha_-$ space (shaded). Overlaid the curve $\sqrt{E}=D_1/\lambda=D_4\lambda$ (red) and curves when further adding non-zero $D_3$  (green).}
\label{fig:NAE01_k0}
\end{figure}
We compute $\textrm{Det}[M_{00, 01, 10, 11}]=0$ to find (compare (\ref{eq:E2233})):
\begin{eqnarray}
\left[\Box_k^3+2\alpha_-\Box_k^2+\left(\beta_E+4E^2\right)\Box_k+\gamma_e+4E^2\alpha_-\right]^2=0,\nonumber\\
\label{eq:E0011}
\end{eqnarray}
and so we may consider the expression inside the square, and treat each solution as a double root.
We again first consider the case $k_x=k_T=0$, for which we have
\begin{eqnarray}
-(\omega^2)^3+2\alpha_+(\omega^2)^2-(\alpha_-^2+7E^2)\omega^2+3E^2\alpha_-=0.\nonumber\\
\end{eqnarray}
Again, we have one real solution and two solutions that are non-real in part of the parameter space, shown in Figure \ref{fig:NAE01_k0} as a function of $\alpha_+$ and $\alpha_-$. Overlaid is the now-familiar curve parametrised by $D_1=\sqrt{E}\lambda$, $D_4=\sqrt{E}/\lambda$, $D_2=D_3=0$ which lies fully in the region of non-real eigenvalues. For $\lambda=1$,  the eigenvalues are $\omega^2=(0, (2\pm i\sqrt{3})E)$. The real part of two eigenvalues is always positive, while the third has negative real part whenever $\alpha_-$ is negative. 

\begin{figure}
\begin{center}
\includegraphics[width=0.35\textwidth]{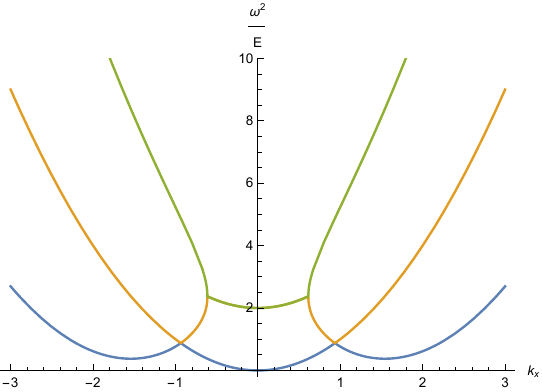}
\includegraphics[width=0.35\textwidth]{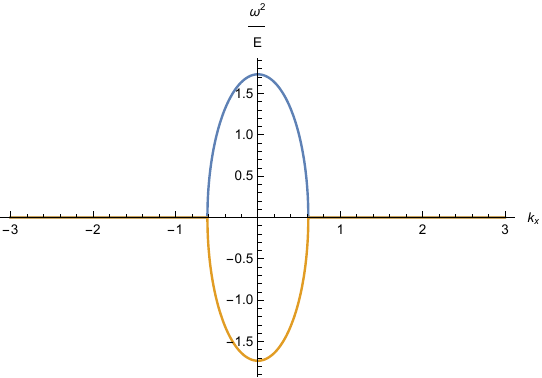}
\includegraphics[width=0.35\textwidth]{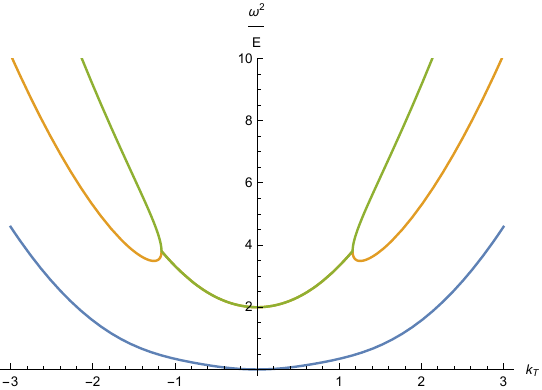}
\includegraphics[width=0.35\textwidth]{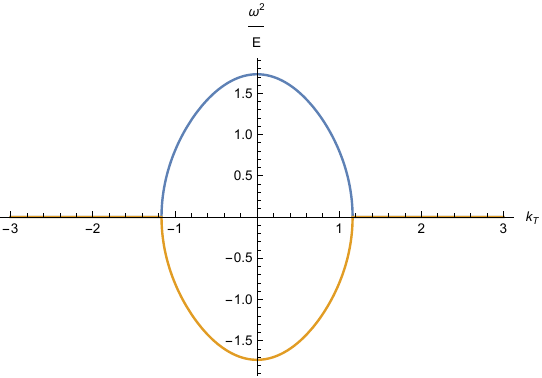}
\end{center}
\caption{Non-Abelian E: Real part and imaginary part for the three (double) eigenvalues in the 0,1 sector, as a function of $k_x$ (top two) and $k_T$ (bottom two), for $k_{T,x}=0$, respectively.}
\label{fig:NAE23_kx_kT}
\end{figure}
For non-zero momentum and $D_1=D_4=\sqrt{E}$, we need to solve
\begin{eqnarray}
&&\left(-\frac{\omega^2}{E}+\frac{k_T^2}{E}+\frac{k_x^2}{E}\right)\times\nonumber\\&&
\qquad\left[
\left(-\frac{\omega^2}{E}+\frac{k_T^2}{E}+\frac{k_x^2}{E}\right)^2-4\left(\frac{\omega^2}{E}+\frac{k_x^2}{E}\right)+3\right]
\nonumber\\&&
\qquad\qquad\qquad-4\left(\frac{\omega^2}{E}-\frac{k_x^2}{E}\right)=0.\nonumber\\
\end{eqnarray}
which provides one real solution and two solutions that are in general non-real. In Figure \ref{fig:NAE23_kx_kT} we show the real and imaginary parts of these eigenvalues as a function of $k_x$ (top) and $k_T$ (bottom), respectively. We see that the eigenvalues are real and positive everywhere except a finite region near the origin (an ellipse in the $k_x-k_T$-plane). 

And so the non-Abelian electric field only, of the set of 12 momentum modes present, several are potentially unstable, depending on the precise choice of $D_{1,..,4}$. For the special case of $D_1=D_4=\sqrt{E}$, the 0,1 sector has a finite momentum region with two unstable modes near the origin with non-real eigenvalues, while in the 2, 3 sector there are two unstable modes in certain momentum regions with real, negative eigenvalues (imaginary $\omega$).

\subsection{Non-Abelian E and B}
\label{sec:NAEB}

\begin{figure}
\begin{center}
\includegraphics[width=0.45\textwidth]{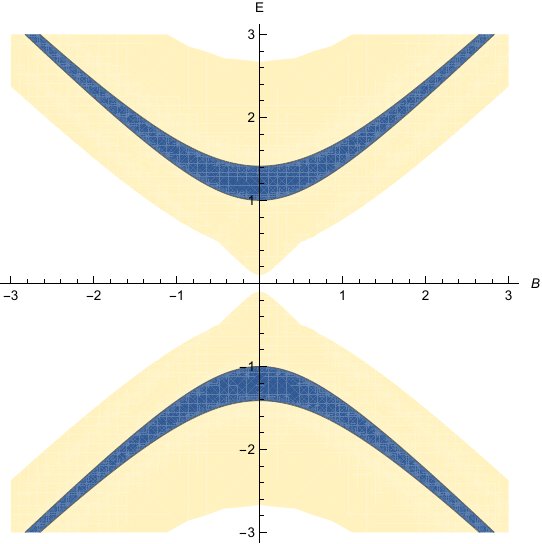}
\includegraphics[width=0.45\textwidth]{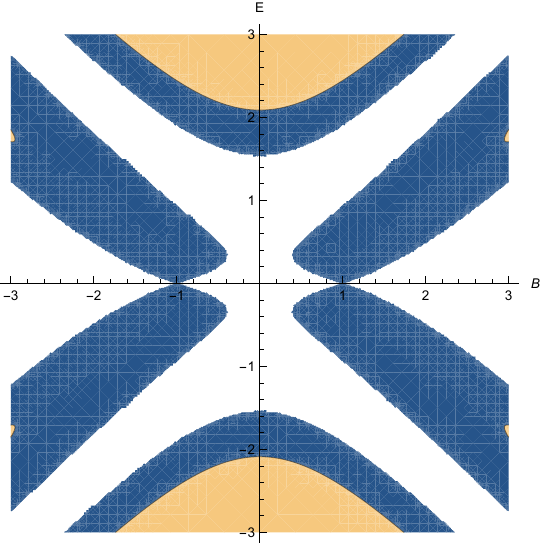}
\end{center}
\caption{Non-Abelian E and B: Regions in the E-B plane, where one or more of the first three (double) eigenvalues has negative real part (top) or non-zero imaginary part (bottom).}
\label{fig:NAEB_k0_contour_1}
\end{figure}

Including both an electric and a magnetic field allows us to specify two directions in space, and hence completely break isotropy.  We recall the non-Abelian/non-Abelian configuration
\begin{eqnarray}
\bA^0_3=D_1,\quad \bA^1_2=D_4,\quad \bA^3_3=D_7,
\end{eqnarray}
which gives
\begin{eqnarray}
E^1_1=gD_1D_4,\quad B^2_1=gD_4D_7.
\end{eqnarray}
The linear equation in momentum space is again a 12-by-12 matrix, but now the sectors are coupled. 
Introducing a shorthand, where we absorb $g$ as before and define
\begin{eqnarray}
&&M_{\rm diag}=\nonumber\\
&&\left(\begin{array}{ccc}
\Box_k-D_1^2+D_4^2+D_7^2&2i\left(D_7k_z-D_1\omega\right)&-2iD_4k_x\\
-2i\left(D_7k_z-D_1\omega\right)&\Box_k-D_1^2+D_7^2&0\\
2iD_4k_x&0&\Box_k+D_4^2
\end{array}\right),\nonumber\\
\end{eqnarray}
and 
\begin{eqnarray}
M_E=
\left(\begin{array}{ccc}
0&0&0\\
0&0&-2E\\
0&2E&0
\end{array}\right),\quad 
M_B=\left(\begin{array}{ccc}
0&0&0\\
0&0&2B\\
0&-2B&0
\end{array}\right),\nonumber\\
\end{eqnarray}
We have schematically
\begin{eqnarray}
\left(\begin{array}{cccc}
M_{\rm diag}&M_E&0&0\\
-M_E^T&M_{\rm diag}&0&M_B\\
0&0&M_{\rm diag}&0\\
0&M_B^T&0&M_{\rm diag}\\
\end{array}\right)
\left(\begin{array}{c}
h^0_{1,2,3}\\
h^1_{1,2,3}\\
h^2_{1,2,3}\\
h^3_{1,2,3}\\
\end{array}\right)=0,\nonumber\\
\end{eqnarray}
This couples the 0,1 and 3 sectors, while the 2-sector is decoupled. Note that this is not simply combining the non-Abelian E and the non-Abelian B that we considered above. 

We write $D_1=E/D_4$ and $D_7=B/D_4$, in which case the determinant of the entire matrix may be conveniently written as
\begin{eqnarray}
&&\frac{1}{D_4^{16}}\left[D_4^4\Box_k^3+2D_4^2\alpha_3\Box_k^2+\beta_3\Box_k+\gamma_3\right]^2\times\nonumber\\
&&\qquad\qquad\left[D_4^4\Box_k^3+2D_4^2\alpha_3\Box_k^2+(\beta_3-4D_4^4(\alpha_3-D_4^4))\Box_k\right.\nonumber\\
&&\qquad\qquad\qquad\qquad\left.+\gamma_3-4D_4^2\alpha_3(\alpha_3-D_4^4)\right]^2=0,\nonumber\\
\end{eqnarray}
with
\begin{eqnarray}
\alpha_3&=&B^2-E^2+D_4^4,\\
\beta_3&=&\alpha_3(\alpha_3+D_4^4)-D_4^8-4D_4^6k_x^2-4D_4^2(Bk_z-E\omega)^2,\nonumber\\\\
\gamma_3&=&D_4^2\left[\alpha_3(\alpha_3-D_4^4)-4D_4^2(\alpha_3-D_4^4)k_x^2\right.\nonumber\\
&&\qquad\qquad\qquad\qquad\left.-4D_4^2(Bk_z-E\omega)^2\right],
\end{eqnarray}
which reduces to (\ref{eq:NAEabg}) in the limit $B=0$ ($D_7=0$). We see that the 12 roots split up into 6 double roots, 3 for each term. 

\begin{figure}
\begin{center}
\includegraphics[width=0.45\textwidth]{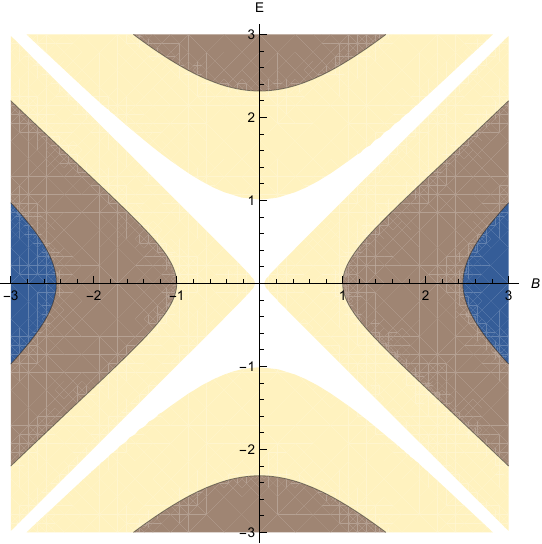}
\includegraphics[width=0.45\textwidth]{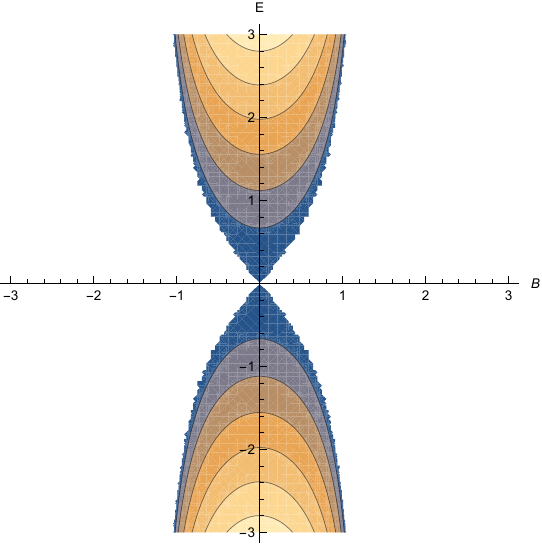}
\end{center}
\caption{Non-Abelian E and B: Regions in the E-B plane, where one or more of the second set of three (double) eigenvalues has negative real part (top) or non-zero imaginary part (bottom).}
\label{fig:NAEB_k0_contour_2}
\end{figure}
We can again specialise to zero momentum, and we will rescale all quantities with the appropriate power of $D_4$ ($\omega/D_4$, $B/D_4^2$, $E/D_4^2$, $\alpha_3/D_4^4$ and so on, which amounts to setting $D_4=1$) to find the requirements
\begin{eqnarray}
\label{eq:EB_eq1}
&&-(\omega^2)^3+2(B^2+E^2+1)(\omega^2)^2\nonumber\\
&&\quad\quad-((B^2-E^2)^2+3B^2+E^2+1)\omega^2\nonumber\\
&&\quad\quad\quad\quad\quad +(B^2-E^2)(B^2-E^2+1)=0,
\end{eqnarray}
and 
\begin{eqnarray}
\label{eq:EB_eq2}
&&-(\omega^2)^3+2(B^2+E^2+1)(\omega^2)^2\nonumber\\
&&\quad\quad-((B^2-E^2)^2-B^2+5E^2+1)\omega^2\nonumber\\
&&\quad\quad\quad\quad\quad -3(B^2-E^2)(B^2-E^2+1)=0,
\end{eqnarray}
These provide 6 distinct (double) eigenvalues for $\omega^2/D_4^2$. 
Figure \ref{fig:NAEB_k0_contour_1} shows the regions in E-B space, where one or more has negative real part (top) or non-zero imaginary part (bottom). Figure \ref{fig:NAEB_k0_contour_2} show the same for the second set of three eigenvalues (solutions of (\ref{eq:EB_eq2})). Ten of twelve eigenmodes are subject to instability in some part of momentum space. 

The solutions for non-zero momentum now involves $k_x$, $k_y$ and $k_z$ independently, in addition to $E$ and $B$, and it is perhaps not so helpful attempting to display the eigenvalues in all generality. But for instance for $k_y\neq 0$, we simply get the same relations but with $\omega^2\rightarrow \omega^2-k_y^2$. Hence all solutions for $\omega^2$ are shifted by $+k_y^2$, so that unstable, but real-$\omega$ solutions become stable for large $k_y$. For $k_x\neq 0$, a similar shift appears, but there are additional terms involving $k_x^2$ as well in both equations
\begin{eqnarray}
+4k_x^2(\omega^2-k_x^2-B^2+E^2).
\end{eqnarray}
Something new happens for $k_z\neq 0$, where in addition to the shift $\omega^2\rightarrow \omega^2-k_z^2$, we pick up terms linear in $\omega$, $k_z$, and proportional to $BE$
\begin{eqnarray}
+4(B^2k_z^2-2BEk_z\omega)(\omega^2-k_z^2-1),
\end{eqnarray}
so that the (relative) signs of $E$, $B$, $k_z$ matter for determining $\omega$. 

We will not pursue this further, but simply note that there is much fun to be had in this large parameter space.

\section{Computing the quadratic and cubic contributions for non-Abelian B}
\label{sec:nonlinear}

\subsection{Setting up the matrix structures: Mode equation}
\label{sec:cubmat}

Having completed our survey of realisations of instabilities, we now return to Eq. (\ref{eq:cubiccont}), which implies that the equation of motion for the fluctuations $h^\mu_a$ including self-interaction at leading order takes the form
\begin{eqnarray}
\left[g^{\mu\nu}(\bar{D}_\rho\bar{D}^\rho)_{ac}+2g \epsilon_{abc}\bar{F}_b^{\mu\nu}\right]h_{\nu,c}-g^2\hat{M}^{\mu\nu}_{ab}(C)h_{\nu,b}=0,\nonumber\\
\end{eqnarray}
where the matrix $\hat{M}^{\mu\nu}_{ab}$ is expressed in terms of the expectation values of the fluctuation fields themselves 
\begin{eqnarray}
C^{\mu,\nu}_{a,b}=\langle h^{\mu}_a(x,t)h^\nu_b(x,t)\rangle=\int \frac{d^3k}{(2\pi)^3}\langle h^{\mu}_a(k,t)(h^\nu_b(k,t))^\dagger\rangle.\nonumber\\
\end{eqnarray}
For the case of a non-Abelian B-field, the matrix structure is as follows (in the 2-3 sector) 
\begin{eqnarray}
\hat{M}^i_{j, ab}=\left[\begin{array}{cccccc}
-\hM_{11}&0&0&0&0&0\\
0&-\hM_{22}&-\hM_{23}&0&0&-\hM_{26}\\
0&-\hM_{32}&-\hM_{33}&0&-\hM_{35}&0\\
0&0&0&-\hM_{44}&0&0\\
0&0&-\hM_{53}&0&-\hM_{55}&-\hM_{56}\\
0&-\hM_{62}&0&0&-\hM_{65}&-\hM_{66}
\end{array}
\right],\nonumber\\
\label{eq:corrmat}
\end{eqnarray}
where 
\begin{eqnarray}
\hM_{11}=C^{3,3}_{2,2}+C^{3,3}_{3,3},\quad
\hM_{22}=C^{3,3}_{3,3}+C^{3,3}_{1,1},\\
\hM_{33}=C^{3,3}_{1,1}+C^{3,3}_{2,2},\quad
\hM_{44}=C^{2,2}_{2,2}+C^{2,2}_{3,3},\\
\hM_{55}=C^{2,2}_{3,3}+C^{2,2}_{1,1},\quad
\hM_{66}=C^{2,2}_{1,1}+C^{2,2}_{2,2},
\end{eqnarray}
and
\begin{eqnarray}
\hM_{23}=\hM_{32}&=-C^{3,3}_{3,2}=-C^{3,3}_{2,3},\\
\hM_{56}=\hM_{65}&=-C^{2,2}_{3,2}=-C^{2,2}_{2,3},\\
\hM_{26}=\hM_{62}&=2C^{2,3}_{2,3}-C^{2,3}_{3,2},\\
\hM_{35}=\hM_{53}&=2C^{2,3}_{3,2}-C^{2,3}_{2,3}.
\end{eqnarray}
We have at this point already implemented that several correlators vanish identically, and we have for simplicity by hand set to zero all correlators involving the modes in the 0-1 sector, since they are stable. The 0-1 sector of the matrix has the same structure as the 2-3 sector, but without the sector-mixing components $\hM_{26,62,35,53}$. 
In our particular setup, we in addition find that $\hM_{11}=\hM_{44}$, $\hM_{22}=\hM_{33}=\hM_{55}=\hM_{66}$, $\hM_{23}=\hM_{32}=\hM_{56}=\hM_{65}$ and $\hM_{35}=\hM_{53}=-\hM_{26}=-\hM_{62}$, as there are only four distinct correlators to take into account, $C^{2,2}_{1,1},C^{2,2}_{2,2},C^{2,3}_{2,3},C^{2,2}_{2,3}$.

Our task is then to compute these time-dependent correlators, and solving again the eigenvalue equations ($\textrm{Det}[\mathcal{M}-g^2\hat{M}]=0$) to find corrected, time-dependent values of $\omega^2$. We will do this below, but already at this stage, we note that in the limit where $\hat{M}\gg \mathcal{M}$, the corrected system now has three double eigenvalues $\hM_{11}, \hM_{22}\pm\sqrt{\hM_{23}^2+\hM_{26}^2}$, of which one is negative whenever $\hM_{23}^2+\hM_{26}^2>\hM_{22}^2$. This means that (some of) the instabilities remain even at late times, provided the background field remains constant. In the 0-1 sector, all modes remain stable, since $\hM_{26}\rightarrow 0$ and as we will see, $\hM_{22}>\hM_{23}$.

\subsection{Setting up the matrix structures: Background equation}
\label{sec:quadmat}

Similarly, we can express the quadratic contribution (\ref{eq:quadraticcont}) in terms of correlators, 
\begin{eqnarray}
\bar{D}^{ab}_\mu \bar{F}^{\mu\nu,b}
=j^{\nu}_a-\bar{M}^{\nu\mu}_{ab}[C]\bA_{\mu,b}.
\end{eqnarray}
Again, in our specific setup the space-derivative terms in the correction to the current $\bar{j}^\nu_a$ vanishes because the correlators are space-independent. The time-derivative terms also vanish because they either act on a stable correlator (in the 0-$\mu$ sector, which we neglect) or because of antisymmetrization. We are therefore left with the matrix $\bar{M}(C)$ which multiplies $\bA$, and we find that this matrix has the same structure as $\hM$. Since there is no original current in the 0-1 sector, it is consistent to set $\bA^{0,1}_a=0$ throughout, and we will consider it no further. In the 2-3 sector, however, we would have to solve the full Yang-Mills equation with a time-dependent current. 

We will not attempt this here, but simply estimate at which time the assumption of constant $\bA$ is likely to break down. We quantify this as when the correction is larger than the original current
\begin{eqnarray}
\bar{M}^{\nu\mu}_{ab}[C]\bA_{\mu,b} > j^\nu_a.
\end{eqnarray}
Recall, that 
\begin{eqnarray}
j^{\nu}_a=-gB
\left[
\left(\begin{array}{c} 0\\0\\0\end{array}\right)
\left(\begin{array}{c} 0\\0\\0\end{array}\right)
\left(\begin{array}{c} 0\\D_7\\-D_8\end{array}\right)
\left(\begin{array}{c} 0\\-D_5\\D_6\end{array}\right)
\right].\nonumber\\
\end{eqnarray}
with $D_6=-D_7=\sqrt{B/g}$, and $D_5=D_8=0$. The correction simply becomes
\begin{eqnarray}
\left(\bar{M}\bA\right)^{\nu}_a=-gB
\left[
\left(\begin{array}{c} 0\\0\\0\end{array}\right)
\left(\begin{array}{c} 0\\0\\0\end{array}\right)
\left(\begin{array}{c} 0\\J_7\\-J_8\end{array}\right)
\left(\begin{array}{c} 0\\-J_5\\J_6\end{array}\right)
\right].\nonumber\\
\end{eqnarray}
with 
\begin{eqnarray}
&&J_6=-J_7=\sqrt{\frac{B}{g}}\frac{g\left(C^{33}_{11}+C^{33}_{33}-3C^{23}_{23}\right)}{B},\nonumber\\
&&J_5=-J_8=\sqrt{\frac{B}{g}}\frac{g\,C^{22}_{23}}{B},
\label{eq:J58}
\end{eqnarray}
so that the form of the total current is unchanged, and will still generate a $B$ field in the $\mu=1$, $a=1$ direction. This $B$-field is however now time dependent, and the time-dependence of $\bar{A}^\mu_a$ could also potentially source an Abelian E-field. This means that as soon as either of the correlators is larger than $B/g$, our approximation breaks down. We will now estimate when that happens. 

\subsection{Eigenvectors, quantization and correlators}

\begin{figure}
\begin{center}
\includegraphics[width=0.45\textwidth]{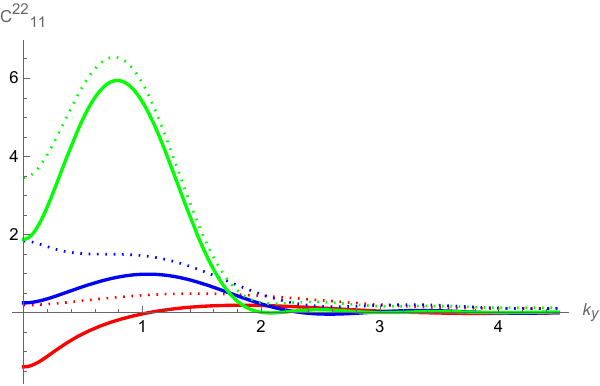}
\includegraphics[width=0.45\textwidth]{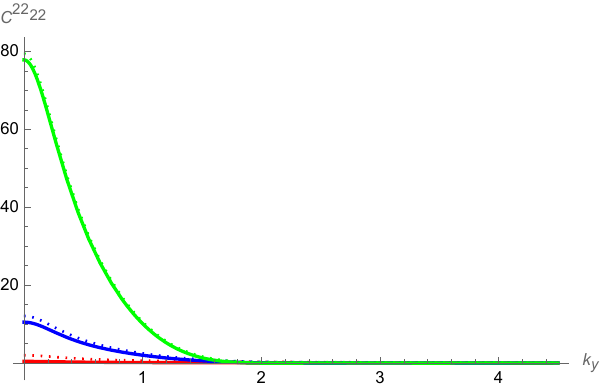}
\includegraphics[width=0.45\textwidth]{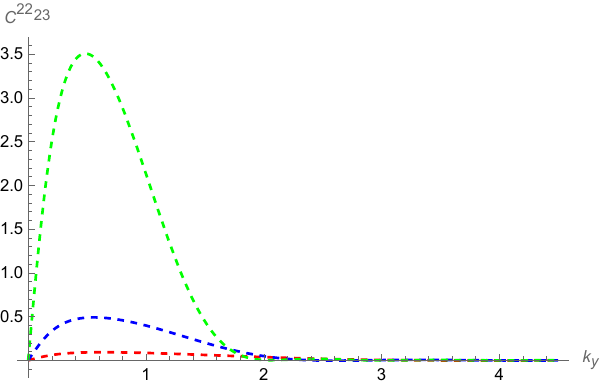}
\includegraphics[width=0.45\textwidth]{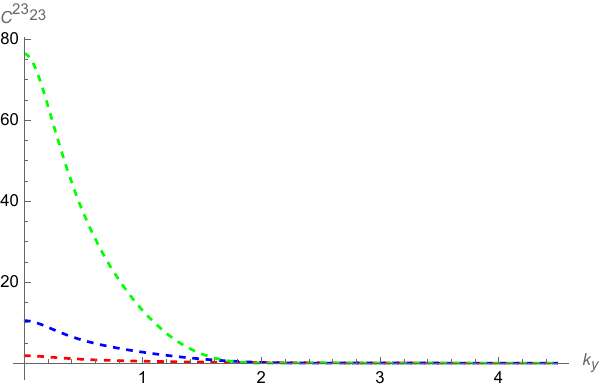}
\end{center}
\caption{Non-Abelian B: The correlators $C^{22}_{11}$, $C^{22}_{22}$, $C^{22}_{23}$ and $C^{23}_{23}$ (top to bottom) as a function of $k_y$ for time $gBt=1,2,3$ (red, blue, green). Renormalised (full lines) and unrenormalised (dotted lines).}
\label{fig:BNA_4corrs}
\end{figure}

The 2-3 sector of the linear system has 6 eigenvalues for each set of momenta $k_{x,y,z}$, $\omega_i^2(k)$, corresponding to 6 eigenvectors $\phi_i(k,t)$ which are solutions to 
\begin{eqnarray}
&&\partial_t^2 \phi_i(k,t)= -\omega_i^2(k)\phi_i(k,t),\quad i=1,...,6,\\
\rightarrow&& \phi_i(k,t) = \alpha_k^ie^{i\omega_i(k)t}+\beta_k^ie^{-i\omega_i(k)t}.
\label{eq:eigensol}
\end{eqnarray}
In order to define the state in which to compute the expectation values, we will assume that the system is prepared in a vanishing background $\bA^\mu_a=0$ for $t<0$, and that $\bA^\mu_a$ is instantaneously turned on at $t=0$, remaining constant for $t>0$. Then we may write for $t<0$
\begin{eqnarray}
\phi_i(k,t) &=& \frac{1}{\sqrt{2\omega^0_k}}\left(a_k^i e^{i\omega_k^0 t}+(a_{-k}^i)^\dagger e^{-i\omega_k^0 t}\right),\\
\partial_t\phi_i(k,t) &=& \frac{i\omega_k^0}{\sqrt{2\omega^0_k}}\left(a_k^i e^{i\omega_k^0 t}-(a_{-k}^i)^\dagger e^{-i\omega_k^0 t}\right),
\end{eqnarray}
where we note that for $\bA^\mu_a=0$, all the eigenvalues coincide $\omega_i(k)\rightarrow \omega^0_k=|{\bf k}|$.

For $t>0$, we may write in general
\begin{eqnarray}
\phi_i(k,t)&=& \alpha_k^i e^{i\omega_k^i t}+\beta_{k}^i e^{-i\omega_k^i t},\\
\partial_t\phi_i(k,t) &=& i\omega_k^i\left(\alpha_k^i e^{i\omega_k^i t}-\beta_{k}^i e^{-i\omega_k^i t}\right).
\end{eqnarray}
Matching the expressions at $t=0$, we can express $\alpha_k^i$ and $\beta_k^i$ in terms of $a_k^i$ and $a_{-k}^i$ as,
\begin{eqnarray}
\alpha_k^i&=&\frac{1}{\sqrt{2\omega^0_k}}\left[a_k^i\left(1+\frac{\omega_k^0}{\omega_k^i}\right)+(a_{-k}^i)^\dagger\left(1-\frac{\omega_k^0}{\omega_k^i}\right)\right],\nonumber\\\\
\beta_k^i&=&\frac{1}{\sqrt{2\omega^0_k}}\left[a_k^i\left(1-\frac{\omega_k^0}{\omega_k^i}\right)+(a_{-k}^i)^\dagger\left(1+\frac{\omega_k^0}{\omega_k^i}\right)\right].\nonumber\\
\label{eq:matching}
\end{eqnarray}
Choosing our initial state as the vacuum at $t<0$, $a_k^i|0\rangle=0$ and using the standard commutation relations $[a_k^i,(a^j_{k'})^\dagger]=(2\pi)^3\delta^{ij}\delta^3(k-k')$, we can straightforwardly compute expectation values of any combination of $\alpha_k^i$, $\beta_k^i$. 

The fluctuation fields $h^{\mu}_a(k)$ in the 2-3 sector are linear combinations of the eigenmodes $\phi_i(k)$, written in terms of a unitary matrix $U^{ij}(k)$. With ($\mu=2,3$, $a=1,2,3$ $\rightarrow$ $1, .., 6$) so that
\begin{eqnarray}
\langle h^i(k) (h^j(k))^\dagger\rangle 
&&=\langle U_{ir}\phi^r(k) [U_{js}\phi^s(k)]^\dagger\rangle\nonumber\\
&&=U^{ir} [U^\dagger]_{rj}\langle \phi^r(k)[\phi^r(k)]^\dagger\rangle,
\label{eq:rotation}
\end{eqnarray}
since $\langle \phi^r(k)[\phi^s(k)]^\dagger\rangle\propto \delta^{rs}$. 
We wish to compute the correlators
\begin{eqnarray}
C^{\mu,\nu}_{a,b}&&=\int \frac{d^3k}{(2\pi)^3}\langle h^{\mu}_a(k)h^{\nu}_b(-k)\rangle\nonumber\\&&
 = \int \frac{d^3k}{(2\pi)^3}\frac{\langle h^i(k)[h^j(k)]^\dagger+[h^i(k)]^\dagger h^j(k)\rangle}{2}.\nonumber\\
\label{eq:fourierback}
\end{eqnarray}
Which follow from combining (\ref{eq:eigensol}, \ref{eq:matching}, \ref{eq:rotation}), and using these we can construct the non-linear correction to the linear mode equation (\ref{eq:corrmat}). 

One final complication is that the diagonal correlators are (IR and UV) divergent, and so in order to gauge the effect of the instability, we will in practice compute
\begin{eqnarray}
&&C^{\mu,\nu}_{a,b;ren}=\nonumber\\&&\int \frac{d^3k}{(2\pi)^3}\bigg[\langle h^{\mu}_a(k,t)h^{\nu}_b(-k,t)\rangle-\langle h^{\mu}_a(k,0)h^{\nu}_b(-k,0)\rangle\bigg].\nonumber\\
\end{eqnarray}
In Figure \ref{fig:BNA_4corrs} we show the four distinct (renormalised) correlators in momentum space at times $t=1,2,3$ (in units where $gB=1$). We see that the $C^{22}_{22}$ and $C^{23}_{23}$ are substantially larger than the other two. 

\subsection{Corrected dispersion relations}

\begin{figure}
\begin{center}
\includegraphics[width=0.45\textwidth]{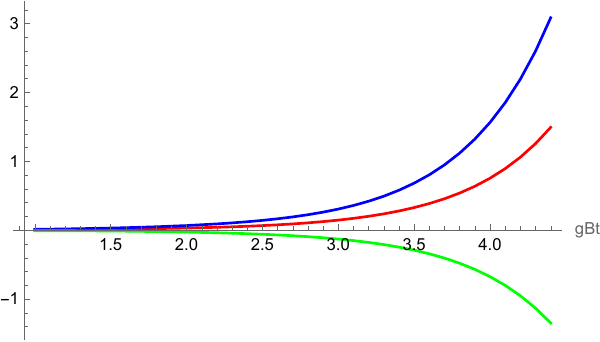}
\end{center}
\caption{Non-Abelian B: The three eigenvalues $M_{11}$, $M_{22}\pm \sqrt{M_23^2-M_{26}^2}$.}
\label{fig:BNA_instabcomp}
\end{figure}

In Figure \ref{fig:BNA_instabcomp}, we show $M_{11}$, $M_{22}\pm\sqrt{M_{23}^2+M_{26}^2}$, which was our late-time estimate for the eigenvalues in section \ref{sec:cubmat}. The lowest eigenvalue is indeed negative throughout, and so for any momentum, for late enough time, instabilities remain if they are not switched off for other reasons. 

\begin{figure}
\begin{center}
\includegraphics[width=0.45\textwidth]{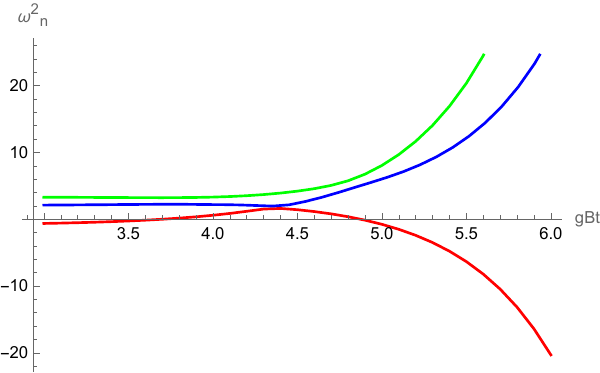}
\end{center}
\caption{Non-Abelian B: The corrected eigenvalues $\omega_i$ as a function of time.}
\label{fig:BNA_corromega}
\end{figure}
We can do better, and simply compute the eigenvalues at $k=0$ of the whole matrix $\mathcal{M}-g^2\hat{M}$, as a function of time. Figure \ref{fig:BNA_corromega} shows the 6 eigenvalues in the 2-3 sector as a function of time. The initially unstable mode is briefly stabilized around $gBt=3.5$, but by $4.7$ the late time behaviour takes over and the eigenvalue becomes negative again.  

\subsection{Corrected current}

\begin{figure}
\begin{center}
\includegraphics[width=0.45\textwidth]{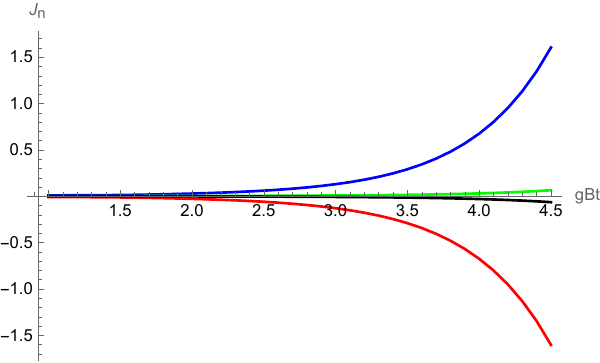}
\end{center}
\caption{Non-Abelian B: The quantities $J_{5,6,7,8}/\sqrt{B/g}$ in time.}
\label{fig:corrcurrents}
\end{figure}
In a similar way, we can insert the correlators into the effective current. Figure \ref{fig:corrcurrents} shows the quantities $J_{5,6,7,8}$ (\ref{eq:J58}) in time. We see that $J_6=-J_7$ grow fastest, and become order 1 (our criterion for $g=B=1$) around $gBt=4-5$. At this time, we can then no longer assume a constant background field. Whether this leads to stability is beyond our approach here, but may be readily studied using numerical simulations.

\section{Conclusions}
\label{sec:conclusion}
To summarize, we have investigated the instabilities of field fluctuations in pure SU(2) classical gauge theory with a constant homogeneous background E and/or B field. We first in section \ref{sec:background} presented a fairly general set of background vector potentials, generalising the setup of \cite{bazak1,bazak2}. We then in section \ref{sec:dispersion} focused on non-Abelian E and B field backgrounds, and derived the dispersion relation that the fluctuations obey. Non-real instances of frequencies $\omega(k)$ signal unstable modes in the linear theory, and we confirmed that these arise for homogeneous E fields, B fields and when combining the two. As an aside, we also found that the instability does not simply depend on the B or E-field, but on the vector field components $\bA^\mu_a$ separately. 

Of particular interest is to understand to what extent these instabilities shut themselves off, due to the non-linearities of the theory. To that end, we computed the non-linear (Gaussian) corrections to both the background field equation and the linear fluctuation equation. We found that the corrections certainly influence the instabilities of the fluctuations, but that rather than turning them off, they can make them stronger. This however assumes that the background field stays constant and homogeneous throughout, which is spoiled by the non-linearities in the background field equation. We estimated the time when the background field is no longer a constant non-Abelian B field to be $gBt\simeq 4$, but further details of how this comes about, and the explicit effect on the instabilities is likely better adressed numerically.

Although the context of this type of instabilities is the thermalisation of the plasma in HICs, anisotropic gauge fields are not unique to QCD. Anisotropic conditions also appear in cosmology, for instance in the context of first order phase transitions. In that case, the walls of nucleated bubbles sweep through the ambient plasma, so that immediately in front and behind the advancing wall, the plasma is anisotropic (see for instance \cite{tranberg,GW1,GW2}). Although it is not clear to what extent net E and B fields are produced, it would be worth investigating whether an anisotropic out-of-equilibrium background could potentially source similar instabilities. 

The obvious continuation of this work is to make a full investigation the other realisations of background E and B fields presented in Appendix \ref{app:4to10}. This could also readily be compared to direct numerical simulations of the pure gauge SU(2) system, going beyond the Gaussian approximation to the non-linearities. Further generalising to SU(3) gauge theory involves writing down a $\mathcal{M}(32\times 32)$ matrix equation and solving for the dispersion relations and then unstable modes there. This is ongoing work, but we note that the results for SU(2) presented here naturally arise as a subsector in the SU(3) theory. 

\section*{Acknowledgments}
We thank Aleksi Kurkela and Germano Nardini for enlightening discussions and Alexander Rothkopf for many useful comments. This work is funded by the grant Project. No. 302640 funded by the Research Council of Norway.

\appendix

\section{Linear equations of motion in momentum space}
\label{sec:dispersion}

If the background field $\bA^\mu_a$ has no space dependence, we can simply Fourier transform in three dimensions and write 
\begin{eqnarray}
h^{\mu}_a(x,y,z,t)= \int \frac{dk_x dk_y dk_z}{(2\pi)^3}e^{-ik_x x-ik_y y-i k_z z}h^{\mu}_a(k_x,k_y,k_z,t).\nonumber\\
\end{eqnarray}
Then the equation of motion in momentum space reads
\begin{eqnarray}
\label{eq:matlin1}
\mathcal{M}(\partial_t^2,\partial_t,k_x,k_y,k_z,\bar{A})h^\mu_a=0,
\end{eqnarray}
which can be solved as a set of coupled linear differential equations. One may further Fourier transform in time, in the sense that
\begin{eqnarray}
h^{\mu}_a(k_x,k_y,k_z,t)=\int \frac{d\omega}{2\pi}h^{\mu}_a(k_x,k_y,k_z,\omega)e^{i\omega t},
\end{eqnarray}
in which case one may substitute $\partial_t\rightarrow i\omega$ in (\ref{eq:matlin1}). The expectation is that for some of the modes (for certain values of $k_x,k_y,k_z, \bar{A}$) $\omega_k$ has a non-zero imaginary part, and the mode grows (or decays) exponentially in time.

If the background field depends on a spatial coordinate, we can only Fourier transform in the remaining coordinates (say $y,z$), to find
\begin{eqnarray}
h^{\mu}_a(x,y,z,t)= \int \frac{dk_y dk_z}{(2\pi)^2}e^{-ik_y y-i k_z z}h^{\mu}_a(x,k_y,k_z,t),\nonumber\\
\end{eqnarray}
and we are led to a relation of the form
\begin{eqnarray}
\label{eq:matlin2}
\mathcal{M}(\partial_t^2,\partial_t,\partial_x^2,\partial_x,k_y,k_z,\bar{A})h^\mu_a=0.
\end{eqnarray}
This implies that the field profile is not a product of plane waves in $x,y,z$ but that the $x$-dependence is some other function to be solved for explicitly.
Similary, if the background field depends on two spatial coordinate, we can only Fourier transform in the remaining coordinates (say, $z$), to find
\begin{eqnarray}
h^{\mu}_a(x,y,z,t)= \int \frac{dk_z}{(2\pi)}e^{-i k_z z}h^{\mu}_a(x,y,k_z,t).
\end{eqnarray}
Neither the $x$- and $y$-dependence can then be expressed as plane waves. Finally, if $\bA^\mu_a$ is a function of $t$, also the Fourier transform in time fails, and the field has a non-trivial time evolution. We again refer to \cite{bazak1} and Appendix \ref{app:4to10} for a discussion on this. 

\section{Space- or time-dependent vector potentials}
\label{app:4to10}

For completeness, we now briefly present a few cases of Abelian vector potentials, and how one might go about solving the linearised equations of motion. 

\subsection{Abelian Configuration of B field ($C_3$ $=$ $B$ $\neq0$) }

Consider a potential $\bar{A}^\mu_a$ in the $\mu=3$ direction with a constant homogeneous magnetic field in the $i=2$ direction and in the $a=1$ color direction,
\begin{equation}
\bar{A}^\mu_a = \begin{bmatrix}
    \begin{pmatrix}
    0 \\
    0 \\
    0
\end{pmatrix}\begin{pmatrix}
    0 \\
    0 \\
    0
\end{pmatrix}\begin{pmatrix}
    0 \\
    0 \\
    0
\end{pmatrix}\begin{pmatrix}
    By \\
    0 \\
    0
\end{pmatrix}
\end{bmatrix}.
\end{equation}
The linearized Y-M equation becomes
\begin{eqnarray}
 \Box h^\mu_a-2gBy\epsilon^{ab1}\partial_zh^\mu_b-2gB\epsilon^{ab1}(\delta^{\mu y}h^z_b-\delta^{\mu z}h^y_b)
\nonumber\\
-g^2B^2y^2\epsilon^{ac1}\epsilon^{cb1}h^\mu_b = 0\label{eqn:magYangMills}.\nonumber\\
\end{eqnarray}
Let us define the functions
\begin{align*}
T^\pm&=h^0_2\pm ih^0_3,         &  X^\pm &=h^x_2\pm ih^x_3,\\
Y^\pm&=h^y_2\pm ih^y_3,       &  Z^\pm&=h^z_2\pm ih^z_3. 
\end{align*}Modifying Eq. (\ref{eqn:magYangMills}) and then solving for each modes in Fourier space, we obtain the dispersion relations such as
\begin{equation}
 \left( -\omega^2+k_x^2+(k_z\mp gBy)^2-\frac{d^2}{dy^2} \right)T^\pm(y) = 0\label{eqn:OsclT},
\end{equation}
\begin{equation}
 \left( -\omega^2+k_x^2+(k_z\mp gBy)^2-\frac{d^2}{dy^2} \right)X^\pm(y) = 0\label{eqn:OsclX},
\end{equation}
\begin{equation}
 \left( -\omega^2\pm2gB+k_x^2+(k_z-gBy)^2-\frac{d^2}{dy^2} \right)U^\pm(y) = 0\label{eqn:OsclU},
\end{equation}
\begin{equation}
 \left( -\omega^2\mp2gB+k_x^2+(k_z+gBy)^2-\frac{d^2}{dy^2} \right)W^\pm(y) = 0\label{eqn:OsclW}.
\end{equation}where
\begin{align*}
U^\pm&\equiv Y^+\pm iZ^+,         &  W^\pm &\equiv Y^-\pm iZ^-.
\end{align*}
The above equations resemble a Schr\"{o}dinger equation for a harmonic oscillator, and have a discrete spectrum of frequencies. We can solve for the frequencies for $W^+$ and $U^-$, to find
\begin{equation}
  \omega_0^2 = 2gB\left(n+\frac{1}{2}\right)+k_x^2.
 \end{equation}and
 \begin{equation}
  \omega_\pm^2 = 2gB\left(n+\frac{1}{2}\right)\pm 2gB+k_x^2.
 \end{equation}
for discrete $n=0,1,2,\ldots$ and continuous $k_x$. 
 \par
 It is clear that $\omega_0^2$ and $\omega_+^2$ are always positive $(\geq 0)$ for any values of $n$, but $\omega_-^2$ is negative whenever $n<\frac{1}{2}-\frac{k_x^2}{2gB}$.
That means there are unstable modes of $U^-$ and $W^+$ which grow exponentially. The remaining modes $T^\pm$ and $X^\pm$, $U^+$ and $W^-$ are stable for all $k_x$. We see that the dependence of $k_z$ is absorbed in a redefinition of the $y$-coordinate. 

\subsection{Abelian Configuration of E field ($C_1$ $=$ $E$ $\neq0$)}

Consider instead a potential $\bar{A}^\mu_a$ in the $\mu=0$ direction with a constant homogeneous electric field in the $i=1$ direction and in the $a=1$ color direction
\begin{equation}
\bar{A}^\mu_a = \begin{bmatrix}
    \begin{pmatrix}
    -Ex \\
    0 \\
    0
\end{pmatrix}\begin{pmatrix}
    0 \\
    0 \\
    0
\end{pmatrix}\begin{pmatrix}
    0 \\
    0 \\
    0
\end{pmatrix}\begin{pmatrix}
    0\\
    0 \\
    0
\end{pmatrix}
\end{bmatrix}.
\end{equation}
The linearized Y-M equation becomes
\begin{eqnarray}
 g^{\mu\nu}\left[ (\partial_\rho\delta^{ad}-gf^{ade}\bar{A}^e_\rho)(\partial^\rho\delta_{dc}-gf_{dcf}\bar{A}^\rho_f)h^c_\nu \right]\nonumber\\
+2gf^{abc}\bar{F}^{\mu\nu}_bh^c_\nu = 0\label{eqn:YME}.\nonumber\\
\end{eqnarray}
Solving Eq. (\ref{eqn:YME}) for the modes in Fourier space give us dispersion relations such as
\begin{equation}
 \left( k_y^2+k_z^2\pm2igE-(\omega+gEx)^2-\frac{d^2}{dx^2} \right)G^\pm(x) = 0\label{eqn:OsclG},
\end{equation}
\begin{equation}
 \left( k_y^2+k_z^2\mp2igE-(\omega-gEx)^2-\frac{d^2}{dx^2} \right)H^\pm(x) = 0\label{eqn:OsclH},
\end{equation}
\begin{equation}
 \left( k_y^2+k_z^2-(\omega\pm gEx)^2-\frac{d^2}{dx^2} \right)Y^\pm(x) = 0\label{eqn:OsclEY},
\end{equation}
\begin{equation}
\left( k_y^2+k_z^2-(\omega\mp gEx)^2-\frac{d^2}{dx^2} \right)Z^\pm(x) = 0\label{eqn:OsclEZ}.
\end{equation}where the functions $G^\pm$ and $H^\pm$ are defined as
\begin{align*}
G^\pm&\equiv T^+\pm X^+,         &  H^\pm &\equiv T^-\pm X^-.
\end{align*}
The above equations again resemble an inverted Schr\"{o}dinger equation, an "upside-down" harmonic oscillator. This does not have any normalizable solutions.

\subsection{Abelian E and Abelian B ($C_1$ $\neq 0$ $\neq$ $C_3$)}

Consider a potential $\bar{A}^\mu_a$ corresponding to a constant homogeneous electric field in the $x$ direction and a constant homogeneous magnetic field in the $y$ direction,
\begin{equation}
\bar{A}^\mu_a = \begin{bmatrix}
    \begin{pmatrix}
    -Ex \\
    0 \\
    0
\end{pmatrix}\begin{pmatrix}
    0 \\
    0 \\
    0
\end{pmatrix}\begin{pmatrix}
    0 \\
    0 \\
    0
\end{pmatrix}\begin{pmatrix}
    By\\
    0 \\
    0
\end{pmatrix}
\end{bmatrix}.
\end{equation}
The linearized Y-M equation in this case becomes
\begin{eqnarray}
 \left( (k_z-gBy)^2-(\omega+gEx)^2\pm2igE-\frac{d^2}{dx^2}-\frac{d^2}{dy^2} \right)\nonumber\\
\times G^\pm(x,y)=0\label{eqn:genAbeOsclEB1G},\nonumber\\
\end{eqnarray}
\begin{eqnarray}
 \left( (k_z+gBy)^2-(\omega-gEx)^2\mp2igE-\frac{d^2}{dx^2}-\frac{d^2}{dy^2} \right)\nonumber\\
\times H^\pm(x,y)=0\label{eqn:genAbeOsclEB1H},\nonumber\\
\end{eqnarray}
\begin{eqnarray}
 \left( (k_z-gBy)^2-(\omega+gEx)^2\pm2gB-\frac{d^2}{dx^2}-\frac{d^2}{dy^2} \right)\nonumber\\
\times U^\pm(x,y)=0\label{eqn:genAbeOsclEB1U},\nonumber\\
\end{eqnarray}
\begin{eqnarray}
 \left( (k_z+gBy)^2-(\omega-gEx)^2\mp2gB-\frac{d^2}{dx^2}-\frac{d^2}{dy^2} \right)\nonumber\\
\times W^\pm(x,y)=0\label{eqn:genAbeOsclEB1W}.\nonumber\\
\end{eqnarray}
Since there is explicit dependence on $x$ and $y$ we are left with a partial differential equation in both coordinates. 
The equations (\ref{eqn:genAbeOsclEB1G}) $-$ (\ref{eqn:genAbeOsclEB1W}) can be further solved variable separation. For example, considering Eq. (\ref{eqn:genAbeOsclEB1U}), we can write
\begin{equation}
 U^\pm(x,y) = U^\pm_E(x)U^\pm_B(y).
\end{equation}
We can split them into two equations such as
\begin{equation}
\left[ Q_U^\pm-\frac{d^2}{dx^2}-g^2E^2\left(\frac{\omega}{gE}+x\right)^2 \right]U^\pm_E(x)=0\label{eqn:invertharmonicEB1},
\end{equation}
\begin{equation}
 \left[ -Q_U^\pm\pm2gB-\frac{d^2}{dy^2}+g^2B^2\left(\frac{k_z}{gB}-y\right)^2 \right]U^\pm_B(y)=0\label{eqn:harmonicEB1}.
\end{equation}where $Q_U^\pm$ is the separation constant.
Equation (\ref{eqn:harmonicEB1}) is another Schr\"{o}dinger equation with the following replacements, $Q_U^\pm\mp2gB$ $\rightarrow$ $2m\varepsilon$, $gB$ $\rightarrow$ $m\bar{\omega}$, $\frac{k_z}{gB}$ $\rightarrow$ $y_0$. Since the oscillator energy $\varepsilon_n$ $=$ $(n+\frac{1}{2})\bar{\omega}$, $n$ $=$ $0,1,2,3,...$
\begin{equation}
 Q_U^\pm = gB(2n+1)\pm2gB.
\end{equation}
Equation (\ref{eqn:invertharmonicEB1}) is an inverted Schr\"{o}dinger equation with the following replacements,
$Q_U^\pm$ $\rightarrow$ $-2m\varepsilon$, $gE$ $\rightarrow$ $m\bar{\omega}$, $\frac{\omega}{gE}$ $\rightarrow$ $x_0$. It has no normalizable solutions. 

Similary, separating variables one can write
\begin{equation}
 W^\pm(x,y) = W^\pm_E(x)W^\pm_B(y).
\end{equation}
It can also be split into two equations such as
\begin{equation}
\left[ Q_W^\pm-\frac{d^2}{dx^2}-g^2E^2\left(\frac{\omega}{gE}-x\right)^2 \right]W^\pm_E(x)=0\label{eqn:invertharmonicEBW},
\end{equation}
\begin{equation}
 \left[ -Q_W^\pm\mp2gB-\frac{d^2}{dy^2}+g^2B^2\left(\frac{k_z}{gB}+y\right)^2 \right]W^\pm_B(y)=0\label{eqn:harmonicEBW}.
\end{equation}where $Q_W^\pm$ is the separation constant.
Similarly one can solve the remaining two equations. 

\subsection{Abelian E and non-Abelian B ($C_2$ $=$ $E$, $D_3$ $=$ $\lambda\sqrt{B/g}$ and $D_{12}$ $=$ $\frac{1}{\lambda}\sqrt{B/g}$)}

Consider a potential $\bar{A}^\mu_a$  corresponding to a constant homogeneous electric field in the $x$ direction and constant homogenous magnetic field components in the $x$ and $z$ directions
\begin{equation}
\bar{A}^\mu_a = \begin{bmatrix}
    \begin{pmatrix}
    0 \\
    0 \\
    0
\end{pmatrix}\begin{pmatrix}
    -tC_2 \\
    0 \\
    D_3
\end{pmatrix}\begin{pmatrix}
    0 \\
    0 \\
    0
\end{pmatrix}\begin{pmatrix}
    D_{12}\\
    0 \\
    0
\end{pmatrix}
\end{bmatrix}.
\end{equation}
In this case, the linearized Yang-Mills equation reduces to
\begin{eqnarray}
&&\Box h^\mu_a+2g\epsilon^{a1b}(D_{12}\partial_z-tC_2\partial_x)h^\mu_b\nonumber\\
&&\quad\quad+2g\epsilon^{a3b}D_3\partial_xh^\mu_b-g^2\epsilon^{a1d}\epsilon^{d1b}
(t^2C_2^2+D_{12}^2)h^\mu_b\nonumber\\
&&\quad\quad+g^2tC_2D_3h^\mu_b(\epsilon^{a1d}\epsilon^{d3b}+\epsilon^{a3d}\epsilon^{d1b})
-g^2\epsilon^{a3d}\epsilon^{d3b}D_3^2h^\mu_b\nonumber\\
&&\quad\quad+2g^2\epsilon^{a2b}(\delta^{\mu3}D_{12}D_3h^x_b-\delta^{\mu1}D_3D_{12}h^z_b) = 0.\nonumber\\
 \label{eqn:ANAgenEB}
\end{eqnarray}
We may now Fourier transform in space, but not in time, and the linear equation for the modes of the fluctuations reduce to a set of 12 coupled linear differential equations which we can split into four $3\times3$ matrices corresponding to the $\mu=0,1,2,3$ sectors introduced earlier, with cross-terms coupling the 0 and 1 sectors. The diagonal $3\times 3$ matrix has the form
\begin{widetext}
\begin{equation}
M^{t}_{ANA} =
 \begin{bmatrix}
 (\Box+g^2D_3^2) & 2igk_xD_3 & g^2tD_3C_2 \\
-2igk_xD_3 & \Box+g^2(t^2C_2^2+D_{12}^2+D_3^2) & 2ig(D_{12}k_z-tC_2k_x) \\
g^2tD_3C_2 & -2ig(D_{12}k_z-tC_2k_x) &\Box+g^2(t^2C_2^2+D_{12}^2)
 \end{bmatrix}
\end{equation}
\end{widetext}
where $\Box$ $=$ $\frac{\partial^2}{\partial t^2}+\textbf{k}^2$, $\textbf{k}^2$ $=$ $k_x^2+k_y^2+k_z^2$. Diagonalizing this set of equations is non-trivial, but would result in a set of 12 non-linear differential equations in $t$, which most likely require numerical evaluation.


\begin{thebibliography}{*}



\bibitem{Mrow3}
S.~Mrowczynski, B.~Schenke and M.~Strickland,
Phys. Rept. \textbf{682} (2017), 1-97
doi:10.1016/j.physrep.2017.03.003
[arXiv:1603.08946 [hep-ph]].

\bibitem{epelbaum1}
T.~Epelbaum and F.~Gelis,
Phys. Rev. Lett. \textbf{111} (2013), 232301
doi:10.1103/PhysRevLett.111.232301
[arXiv:1307.2214 [hep-ph]].

\bibitem{epelbaum2}
T.~Epelbaum and F.~Gelis,
Phys. Rev. D \textbf{88} (2013), 085015
doi:10.1103/PhysRevD.88.085015
[arXiv:1307.1765 [hep-ph]].

\bibitem{lappi1}
T.~Lappi,
Phys. Rev. C \textbf{67} (2003), 054903
doi:10.1103/PhysRevC.67.054903
[arXiv:hep-ph/0303076 [hep-ph]].

\bibitem{lappi2}
T.~Lappi and L.~McLerran,
Nucl. Phys. A \textbf{772} (2006), 200-212
doi:10.1016/j.nuclphysa.2006.04.001
[arXiv:hep-ph/0602189 [hep-ph]].

\bibitem{lappi3}
F.~Gelis, K.~Kajantie and T.~Lappi,
Phys. Rev. Lett. \textbf{96} (2006), 032304
doi:10.1103/PhysRevLett.96.032304
[arXiv:hep-ph/0508229 [hep-ph]].

\bibitem{Baier}
R.~Baier, A.~H.~Mueller, D.~Schiff and D.~T.~Son,
Phys. Lett. B \textbf{502} (2001), 51-58
doi:10.1016/S0370-2693(01)00191-5
[arXiv:hep-ph/0009237 [hep-ph]].

\bibitem{Mrow1}
S.~Mrowczynski,
Phys. Lett. B \textbf{314} (1993), 118-121
doi:10.1016/0370-2693(93)91330-P

\bibitem{Mrow2}
S.~Mrowczynski,
Acta Phys. Polon. B \textbf{37} (2006), 427-454
[arXiv:hep-ph/0511052 [hep-ph]].


\bibitem{kurk1}
A.~Kurkela and G.~D.~Moore,
JHEP \textbf{12} (2011), 044
doi:10.1007/JHEP12(2011)044
[arXiv:1107.5050 [hep-ph]].

\bibitem{kurk2}
A.~Kurkela and G.~D.~Moore,
JHEP \textbf{11} (2011), 120
doi:10.1007/JHEP11(2011)120
[arXiv:1108.4684 [hep-ph]].


\bibitem{kin1}
J.~Berges, K.~Boguslavski, S.~Schlichting and R.~Venugopalan,
Phys. Rev. D \textbf{89} (2014) no.7, 074011
doi:10.1103/PhysRevD.89.074011
[arXiv:1303.5650 [hep-ph]].

\bibitem{kin2}
A.~Kurkela, A.~Mazeliauskas, J.~F.~Paquet, S.~Schlichting and D.~Teaney,
Phys. Rev. Lett. \textbf{122} (2019) no.12, 122302
doi:10.1103/PhysRevLett.122.122302
[arXiv:1805.01604 [hep-ph]].

\bibitem{kin3}
A.~Kurkela, A.~Mazeliauskas, J.~F.~Paquet, S.~Schlichting and D.~Teaney,
Phys. Rev. C \textbf{99} (2019) no.3, 034910
doi:10.1103/PhysRevC.99.034910
[arXiv:1805.00961 [hep-ph]].

\bibitem{kin4}
A.~Kurkela and Y.~Zhu,
Phys. Rev. Lett. \textbf{115} (2015) no.18, 182301
doi:10.1103/PhysRevLett.115.182301
[arXiv:1506.06647 [hep-ph]].

\bibitem{schlichting1}
J.~Berges and S.~Schlichting,
Phys. Rev. D \textbf{87} (2013) no.1, 014026
doi:10.1103/PhysRevD.87.014026
[arXiv:1209.0817 [hep-ph]].

\bibitem{schlichting2}
J.~Berges, S.~Scheffler, S.~Schlichting and D.~Sexty,
Phys. Rev. D \textbf{85} (2012), 034507
doi:10.1103/PhysRevD.85.034507
[arXiv:1111.2751 [hep-ph]].

\bibitem{schlichting3}
J.~Berges, K.~Boguslavski, S.~Schlichting and R.~Venugopalan,
Phys. Rev. D \textbf{89} (2014) no.7, 074011
doi:10.1103/PhysRevD.89.074011
[arXiv:1303.5650 [hep-ph]].

\bibitem{Randrup}
J.~Randrup and S.~Mrowczynski,
Phys. Rev. C \textbf{68} (2003), 034909
doi:10.1103/PhysRevC.68.034909
[arXiv:nucl-th/0303021 [nucl-th]].


\bibitem{Arnold}
P.~B.~Arnold and J.~Lenaghan,
Phys. Rev. D \textbf{70} (2004), 114007
doi:10.1103/PhysRevD.70.114007
[arXiv:hep-ph/0408052 [hep-ph]].


\bibitem{sim1}
P.~B.~Arnold, J.~Lenaghan, G.~D.~Moore and L.~G.~Yaffe,
Phys. Rev. Lett. \textbf{94} (2005), 072302
doi:10.1103/PhysRevLett.94.072302
[arXiv:nucl-th/0409068 [nucl-th]].

\bibitem{sim2}
A.~Rebhan, P.~Romatschke and M.~Strickland,
Phys. Rev. Lett. \textbf{94} (2005), 102303
doi:10.1103/PhysRevLett.94.102303
[arXiv:hep-ph/0412016 [hep-ph]].

\bibitem{sim3}
P.~B.~Arnold, G.~D.~Moore and L.~G.~Yaffe,
Phys. Rev. D \textbf{72} (2005), 054003
doi:10.1103/PhysRevD.72.054003
[arXiv:hep-ph/0505212 [hep-ph]].

\bibitem{sim4}
P.~Romatschke and A.~Rebhan,
Phys. Rev. Lett. \textbf{97} (2006), 252301
doi:10.1103/PhysRevLett.97.252301
[arXiv:hep-ph/0605064 [hep-ph]].

\bibitem{sim5}
D.~Bodeker and K.~Rummukainen,
JHEP \textbf{07} (2007), 022
doi:10.1088/1126-6708/2007/07/022
[arXiv:0705.0180 [hep-ph]].

\bibitem{bazak1}
S.~Bazak and S.~Mrowczynski,
Phys. Rev. D \textbf{105} (2022) no.3, 034023
doi:10.1103/PhysRevD.105.034023
[arXiv:2111.11396 [hep-ph]].

\bibitem{bazak2}
S.~Bazak and S.~Mrowczynski,
Phys. Rev. D \textbf{106} (2022) no.3, 034031
doi:10.1103/PhysRevD.106.034031
[arXiv:2205.08282 [hep-ph]].

\bibitem{tranberg}
Z.~G.~Mou, P.~M.~Saffin and A.~Tranberg,
JHEP \textbf{02} (2021), 189
doi:10.1007/JHEP02(2021)189
[arXiv:2006.13620 [hep-th]].

\bibitem{GW1}
M.~Hindmarsh, S.~J.~Huber, K.~Rummukainen and D.~J.~Weir,
Phys. Rev. Lett. \textbf{112} (2014), 041301
doi:10.1103/PhysRevLett.112.041301
[arXiv:1304.2433 [hep-ph]].

\bibitem{GW2}
C.~Caprini, M.~Chala, G.~C.~Dorsch, M.~Hindmarsh, S.~J.~Huber, T.~Konstandin, J.~Kozaczuk, G.~Nardini, J.~M.~No and K.~Rummukainen, \textit{et al.}
JCAP \textbf{03} (2020), 024
doi:10.1088/1475-7516/2020/03/024
[arXiv:1910.13125 [astro-ph.CO]].


\end{thebibliography}
\end{document}